\pdfoutput=1
\documentclass[11pt]{article}

\usepackage[final]{acl}

\usepackage{times}
\usepackage{latexsym}
\usepackage[utf8]{inputenc} % allow utf-8 input
\usepackage[T1]{fontenc}    % use 8-bit T1 fonts
\usepackage{hyperref}       % hyperlinks
\usepackage{url}            % simple URL typesetting
\usepackage{booktabs}       % professional-quality tables
\usepackage{amsfonts}       % blackboard math symbols
\usepackage{nicefrac}       % compact symbols for 1/2, etc.
\usepackage{microtype}      % microtypography
\usepackage{xcolor}         % colors
\usepackage{graphicx}
\usepackage{amsmath}
\usepackage{amsfonts}
\usepackage{multirow}
\usepackage{subfigure}
\usepackage{subcaption}
\usepackage{pgfplots}
\usepackage{xspace}
\usepackage{enumitem}
\usepackage{wrapfig}
\usepackage{bbding}
\usepackage{colortbl}
\usepackage{adjustbox}
\usepackage{float}
\usepackage[T1]{fontenc}
\usepackage[utf8]{inputenc}

\usepackage{microtype}

\usepackage{inconsolata}

\usepackage{graphicx}

\definecolor{mycolor_green}{HTML}{D5E8D4}
\definecolor{mycolor_orange}{HTML}{FFE6CC}
\definecolor{mycolor_blue}{HTML}{DAE8FC}
\definecolor{mycolor_red}{HTML}{F8CECC}

\definecolor{mycolor_red1}{HTML}{ea3323}
\definecolor{mycolor_red2}{HTML}{ef685c}
\definecolor{mycolor_red3}{HTML}{f49991}

\definecolor{mycolor_green1}{HTML}{9fd77f}
\definecolor{mycolor_green2}{HTML}{b8e2a0}
\definecolor{mycolor_green3}{HTML}{def1d3}

\usepackage{mdframed}

\newcommand*\samethanks[1][\value{footnote}]{\footnotemark[#1]}

\title{AdaSteer: Your Aligned LLM is Inherently an Adaptive Jailbreak Defender}

\author{Weixiang Zhao$^1$\thanks{\ \ \ Equal contribution}, Jiahe Guo$^1$\samethanks, Yulin Hu$^1$, Yang Deng$^2$, An Zhang$^3$, Xingyu Sui$^1$ \\ \textbf{Xinyang Han}$^1$, \textbf{Yanyan Zhao}$^1$\thanks{\ \ Corresponding author}, \textbf{Bing Qin}$^1$, \textbf{Tat-Seng Chua}$^3$, \textbf{Ting Liu}$^1$ \\
        $^1$Harbin Institute of Technology,
        $^2$Singapore Management University \\ $^3$National University of Singapore\\
        \texttt{\{wxzhao, jhguo, yyzhao\}@ir.hit.edu.cn}}

\begin{document}
\maketitle
\begin{abstract}
Despite extensive efforts in safety alignment, large language models (LLMs) remain vulnerable to jailbreak attacks. Activation steering offers a training-free defense method but relies on fixed steering coefficients, resulting in suboptimal protection and increased false rejections of benign inputs. To address this, we propose AdaSteer, an adaptive activation steering method that dynamically adjusts model behavior based on input characteristics. We identify two key properties: Rejection Law (R-Law), which shows that stronger steering is needed for jailbreak inputs opposing the rejection direction, and Harmfulness Law (H-Law), which differentiates adversarial and benign inputs. AdaSteer steers input representations along both the Rejection Direction (RD) and Harmfulness Direction (HD), with adaptive coefficients learned via logistic regression, ensuring robust jailbreak defense while preserving benign input handling. Experiments on LLaMA-3.1, Gemma-2, and Qwen2.5 show that AdaSteer outperforms baseline methods across multiple jailbreak attacks with minimal impact on utility. Our results highlight the potential of interpretable model internals for real-time, flexible safety enforcement in LLMs. Our code is available at: \href{https://github.com/MuyuenLP/AdaSteer}{https://github.com/MuyuenLP/AdaSteer}. \textcolor{red}{WARNING: This paper may contain content that is offensive and harmful.}
\end{abstract}

\section{Introduction}
Despite extensive efforts have been made for safety alignment of large language models (LLMs) \citep{ouyang2022training, bai2022constitutional, askell2021general}, studies show that even well-aligned models remain vulnerable to jailbreak attacks, where adversarial prompts successfully bypass their safety mechanisms \citep{wei2024jailbroken, jones2023automatically, zou2023universal, carlini2024aligned}.
The prevailing defense strategy against such vulnerabilities is safety post-training, where models undergo additional fine-tuning on curated safety data to reinforce their safeguards. However, this approach is computationally expensive \citep{zaremba2025trading} and highly dependent on the quality and diversity of the training dataset \citep{wang2024data}, leading to significant variability in efficacy.

\begin{figure*}[t]
    \centering
    \includegraphics[width=1.0\textwidth]{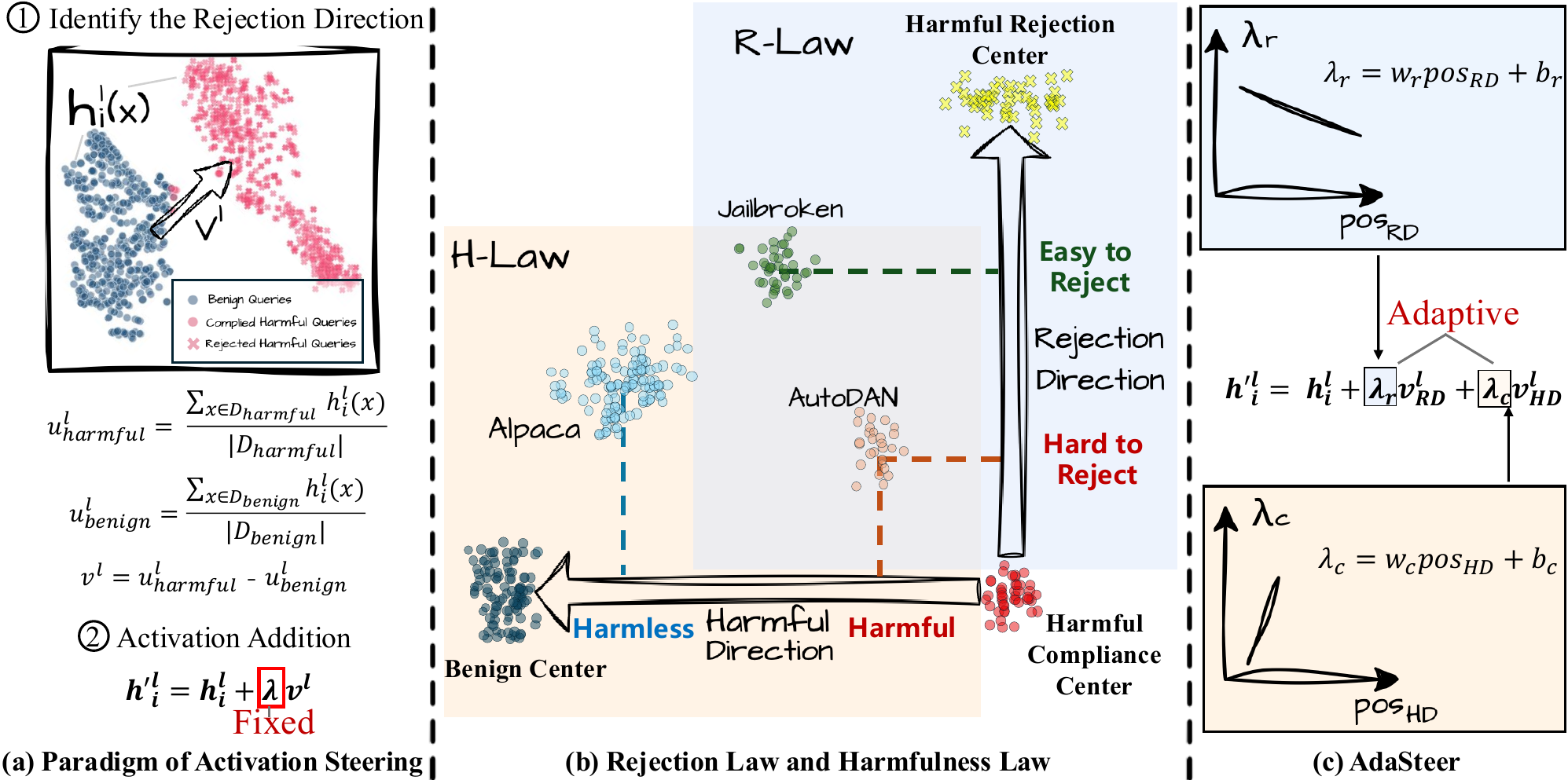}
    \caption{The overall comparison between previous activation steering and 
    our AdaSteer. (a) The two-step paradigm of activation steering, with the fixed steering coefficient $\lambda$. (b) Deriving rejection law and harmfulness law. (c) We propose \textbf{AdaSteer} to achieve real-time, adaptive and input-dependent jailbreak defense.}
    \label{fig:Intro}
    \vspace{-3mm}
\end{figure*}

Activation steering offers a promising training-free alternative by directly manipulating a model’s internal representations along the rejection direction within its activation space \citep{turner2023activation,zou2023representation,panickssery2023steering,arditi2024refusal}. This technique is grounded in the theoretical premise that LLMs encode features or concepts as linear directions in activation space \citep{mikolov2013linguistic,park2024linear}. As illustrated in Figure \ref{fig:Intro}(a), at the model layer $l$, this method first identifies the model’s intrinsic rejection direction with representations of benign and harmful inputs, and extract a rejection steering vector, represented as $\boldsymbol{v}^l$. During inference, a simple activation addition step is performed with a \emph{fixed} strength scalar $\lambda$, steering the input representation toward the rejection region.

However, existing activation steering methods suffer from a key limitation: they lack \emph{dynamic} adaptation to varying input contexts. The fixed steering coefficient $\lambda$ is applied indiscriminately across all inputs, leading to two major challenges: (1) for jailbreak inputs, different attack strategies exhibit diverse characteristics, meaning that applying a static steering coefficient $\lambda$ often results in suboptimal protection \citep{stickland2024steering,shen2025jailbreak,lee2025programming}; (2) for benign inputs, such reinforcement of refusal behavior significantly increases the risk of false rejections, limiting the model’s overall utility \citep{qian2024hsf,bhattacharjee2024towards,arditi2024refusal}. These issues highlight the need for \emph{an adaptive activation steering mechanism that can dynamically adjust its strength based on input characteristics.}

Inspired by recent interpretability studies \citep{leong2024no,zheng2024prompt,zhang2025jbshield} suggesting that LLM rejection behaviors are governed by two key factors: (1) assessing input harmfulness and (2) deciding whether to reject, we seek to perform a dual-direction steering that adjusts model activations along both the \emph{Rejection Direction} (RD) and the \emph{Harmfulness Direction} (HD).

To address the first challenge, we conduct an empirical analysis of different types of jailbreak inputs along the RD within three safety-aligned LLMs: LLaMA-3.1 \citep{dubey2024llama}, Gemma-2 \citep{team2024gemma}, and Qwen2.5 \citep{yang2024qwen2}. As shown in Figure \ref{fig:Intro}(b), we identity RD using contrastive pairs of complied (red cluster) and rejected (yellow cluster) harmful instructions via the difference-in-means technique \citep{belrose2023diff}. We surprisingly find that different jailbreak types exhibit distinct patterns along RD, which can be summarizd as the Rejection Law (R-Law):
\begin{mdframed}[backgroundcolor=mycolor_blue!50, hidealllines=true]
\textbf{Rejection Law}: Along RD, jailbreak types that are positioned further against the rejection direction are more difficult for the backbone model to defend against.
\end{mdframed}
Thus, R-Law can be leveraged as: the farther an input is along RD against the rejection direction, (i.e., the more adversary it is), the stronger \textbf{rejection steering} should be applied to enforce rejection.

%This pattern is well-approximated by a simple linear regression, indicating that the more distant inputs are from harmful compliance center, the stronger refusal behavior should be reinforced.

However, solely depending on R-Law can not solve the second challenge as benign inputs can sometimes also exhibit distributions that oppose the rejection direction along RD, making them appear similar to jailbreak inputs. This directly motivates us to identity and leverage HD, reflecting the harmfulness of different inputs accordingly. Similarly, we obtain HD by contrasting complied harmful instructions with benign ones (blue cluster) and Harmfulness Law (H-Law) is derived:
\begin{mdframed}[backgroundcolor=mycolor_orange!50, hidealllines=true]
\textbf{Harmfulness Law}: Along HD, jailbreak inputs shift further toward harmfulness compared to benign inputs (blue cluster), confirming their harmful nature and distinguishing them from benign queries.
\end{mdframed}
Since HD represents the backbone’s compliance behavior---identified by benign and harmful inputs that are both complied by the model---H-Law can be interpreted and leveraged as follows: the farther an input is along HD against the harmfulness direction, (i.e., the safer it is), the stronger the \textbf{compliance steering} should be applied along HD.

Building on these critical insights, we propose a novel dual-direction \textbf{\underline{Ada}}ptive activation \textbf{\underline{Steer}}ing method for jailbreak defense (\textbf{AdaSteer}), enabling dynamic and input-dependent control. As illustrated in Figure \ref{fig:Intro}(c), AdaSteer steers the input representation using two steering vectors, $\boldsymbol{v}^l_{\text{RD}}$ and $\boldsymbol{v}^l_{\text{HD}}$, along the Rejection Direction (RD) and Harmfulness Direction (HD), respectively. The corresponding coefficients,  $\lambda_r$ and $\lambda_c$, are determined via logistic regression based on the Rejection Law (R-Law) and Harmfulness Law (H-Law). For jailbreak inputs, AdaSteer dynamically adjusts $\lambda_r$ to reinforce rejection while keeping $\lambda_c$ minimal to prevent interference. For benign inputs, a larger $\lambda_c$ is applied, steering the representation toward compliance behavior and preserving model utility.

It is important to emphasize that the direction identification and logistic regression fitting process relies solely on standard harmful prompts, with only a small development set of jailbreak data used for adjustment. This set has no overlap with the final test data, ensuring a fair evaluation. This highlights that our AdaSteer enables \textbf{real-time} and \textbf{flexible} safety enforcement, dynamically adapting to emerging attack strategies. As a result, it represents an \textbf{adaptive defense} mechanism that merits further exploration \citep{anthropic2025blog}.

Experiments on LLaMA-3.1-8B-Instruct \citep{dubey2024llama}, Gemma-2-9B-it \citep{team2024gemma}, and Qwen2.5-7B-Instruct \citep{yang2024qwen2} validate that R-Law and H-Law hold broadly. AdaSteer consistently outperforms baseline methods in jailbreak defense across 7 attack strategies. Furthermore, AdaSteer minimally affects the model’s performance on benign inputs, ensuring its utility remains intact. Our work serves as a concrete demonstration that insights gained from interpreting model internals can have practical applications and well-aligned LLMs hold significant potential to function as adaptive jailbreak defenders.

\section{Preliminaries}

\paragraph{Jailbreak Attacks and Defenses}
A \emph{jailbreak attack} seeks to craft an adversarial prompt $s' = \mathcal{A}(s_0)$, where $\mathcal{A}$ represents an attack method and $s_0$ is a vanilla harmful prompt. The objective is to induce the LLM to generate a harmful response that aligns with the malicious intent of  $s_0$, bypassing built-in safety mechanisms. Conversely, a \emph{jailbreak defense} aims to protect the model against such adversarial manipulations.

\paragraph{Activation Steering} Existing research suggests that LLMs encode features or concepts as linear directions in activation space \citep{mikolov2013linguistic, park2024linear}. Building on this insight, activation steering aims to directly control model behavior by adjusting its internal activations along specific feature directions during inference. This method generally follows two key steps. First, at the specific model layer $l$, a steering vector $\boldsymbol{v}^l$ is derived along the desired feature direction, typically by computing the difference in activations between examples that exhibit the target behavior and those that do not. Second, during inference, this vector is introduced into the model’s hidden states $h^l_i$ at the $i$-th token position within the selected layer $l$, scaled by a coefficient $\lambda$:
\begin{equation*}
     \boldsymbol{h'}^{l}_i = \boldsymbol{h}^{l}_i + \lambda \, \boldsymbol{v}^l
\end{equation*}
where $i$ represents the index of the token’s representation in the input, while $l$ denotes the index of the manipulated layer. %$\lambda$ is the coefficient.

\section{Methodology}
\subsection{Overview}
We propose AdaSteer, which dynamically steers the model’s activations based on the input’s characteristics, ensuring strong resistance against adversarial prompts while minimizing unnecessary refusals of benign queries. The adaptive steering mechanism is formulated as follows:
\begin{equation}\label{eq:adasteer}
\boldsymbol{h’}_{i}^l = \boldsymbol{h}^{l}_i + \lambda_r \boldsymbol{v}^l_{\text{RD}} + \lambda_c \boldsymbol{v}^l_{\text{HD}}
\end{equation}
where RD (Rejection Direction) and HD (Harmfulness Direction) represent key axes within the activation space that encode the model’s refusal and harmfulness behaviors, respectively. The corresponding steering vectors $\boldsymbol{v}^l_{\text{RD}}$ and $\boldsymbol{v}^l_{\text{HD}}$ adjust the model’s activations, with their strengths  $\lambda_r$ and $\lambda_c$ dynamically determined using logistic regression. The following sections introduce how we identify these directions, extract steering vectors, and determine the adaptive coefficients.

\subsection{Rejection Direction (RD), $\boldsymbol{v}_{\text{RD}}$ and $\lambda_r$}
LLMs encode rejection behaviors as a linear direction within the activation space \citep{arditi2024refusal}. We identify this Rejection Direction (RD) and analyze how different jailbreak strategies exhibit distinct behaviors along it, laying the foundation for an adaptive rejection mechanism through input-dependent steering strength $(\lambda_r)$.

\paragraph{Datasets} We utilize two types of vanilla harmful data to identify RD---one consisting of inputs rejected by the model and the other containing those that bypassed rejection. These harmful samples are sourced from multiple datasets, including AdvBench \citep{zou2023universal}, TDC2023 \citep{mazeika2023trojan, mazeika2024harmbench}, Malicious Instruct \citep{huang2024catastrophic}, and Jailbreak Bench \citep{chao2024jailbreakbench}.

\paragraph{Identifying RD} To identify RD, we compute the difference between the model’s mean activations when processing \emph{rejected} and \emph{complied} harmful inputs. This approach, known as the difference-in-means method \citep{belrose2023diff}, effectively isolates the RD by capturing activation shifts associated with rejection behavior. For each layer $l \in [L]$, we calculate the mean activation $\boldsymbol{\mu}^{l}_{\text{\scriptsize r-harmful}}$ for rejected harmful inputs from $D^{\text{\tiny rejection}}_{\text{\tiny harmful}}$ and $\boldsymbol{\mu}^{l}_{\text{\scriptsize c-harmful}}$ for complied harmful inputs from $D^{\text{\tiny compliance}}_{\text{\tiny  harmful}}$, with the representation of the last token position $\boldsymbol{h}^{l}(x)$ given the input $x$:
\begin{align}
\boldsymbol{\mu}^{l}_{\text{\scriptsize r-harmful}} &= \frac{1}{|D^{\text{\tiny  rejection}}_{\text{\tiny  harmful}}|} \sum\nolimits_{x \in D^{\text{\tiny  rejection}}_{\text{\tiny  harmful}}} \boldsymbol{h}^{l}(x) \\
\boldsymbol{\mu}^{l}_{\text{\scriptsize c-harmful}} &= \frac{1}{|D^{\text{\tiny compliance}}_{\text{\tiny harmful}}|} \sum\nolimits_{x \in D^{\text{\tiny compliance}}_{\text{\tiny harmful}}} \boldsymbol{h}^{l}(x)
\end{align}
We then identity RD via difference-in-means:
\begin{align}
\label{equ:RD}
\boldsymbol{d}^{l}_{\text{RD}} &= \boldsymbol{\mu}^{l}_{\text{\scriptsize r-harmful}} - \boldsymbol{\mu}^{l}_{\text{\scriptsize c-harmful}}
\end{align}

\paragraph{Extracting Rejection Steering Vector} Unlike prior works that conducts extensive search and validation to identify the most salient direction \citep{arditi2024refusal,shen2025jailbreak}, we directly use $\boldsymbol{d}^{l}_{\text{RD}}$ as the steering vector $\boldsymbol{v}_{\text{RD}}^{l}$ at each layer and each token position, which still exhibits significant effects on steering rejection behavior.

\begin{figure}
  \centering
  \includegraphics[width=\linewidth]{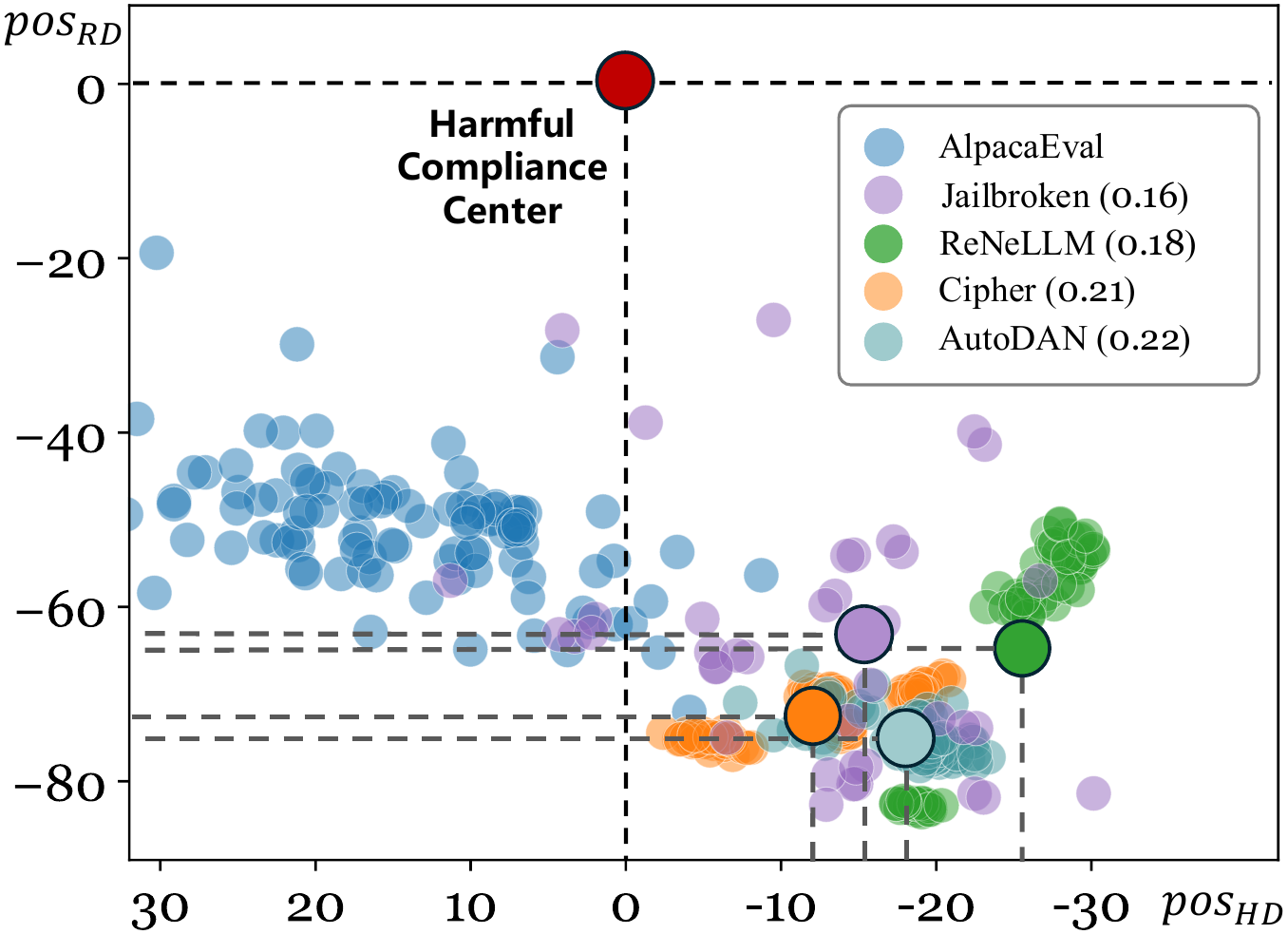}
  \caption{The visualization of $pos_{\text{RD}}$ and $pos_{\text{HD}}$ for each input. The value in parentheses next to each jailbreak method in the legend indicates the average $\lambda_r$ needed to cause the model to reject all inputs.}
  \label{fig:llama31_dis}
\end{figure}

\paragraph{Deriving the Rejection Law}
As illustrated in Figure \ref{fig:llama31_dis}, jailbreak inputs exhibit distinct distributions along RD. We define the Harmful Compliance Center (red point) as the origin, where positive values correspond to increased rejection and negative values indicate compliance tendencies. We observe an almost linear relationship between an input’s RD position ($pos_{\text{RD}}$) and the required rejection steering strength ($\lambda_r$), which forms the Rejection Law:

\begin{mdframed}[backgroundcolor=mycolor_blue!50, hidealllines=true]
\textbf{Rejection Law}: Inputs that are positioned further in the negative direction against RD require a greater rejection steering coefficient $\lambda_r$ to induce rejection behavior.
\end{mdframed}

\paragraph{Fitting the Rejection Law} Formally, $pos_{\text{RD}}$ can be obtained by:
\begin{align}
\label{equ:dis_rd}
pos_{\text{RD}} &=   (\boldsymbol{h}^{l}-\boldsymbol{\mu}^{l}_{\text{\scriptsize c-harmful}}) \cdot \boldsymbol{d}_{\text{RD}}^{l}
\end{align}
We adopt those harmful inputs that make the backbone comply, apply steering with varying strengths $\lambda_r$, and record both the original $pos_{\text{RD}}$ of each harmful input and the corresponding $\lambda_r$ used to induce rejection behavior, forming ($pos_{\text{RD}}$, $\lambda_r$) pairs. Then we fit a logistic regression curve:
\begin{equation}
\label{eq:linear_r}
\lambda_{r} = w_r \cdot {pos_{\text{RD}}} + b_r
\end{equation}
where $w_r$, $b_r$ are hyperparameters in logistic regression. We conduct a grid search on the validation set to fine-tune the curve with greater precision.

\subsection{Harmfulness Direction (HD), $\boldsymbol{v}_{\text{HD}}$ and $\lambda_c$}
Relying solely on RD can lead to false rejections of benign inputs, as they may also distribute negatively along RD. To address this, we introduce the Harmfulness Direction (HD), capturing harmfulness characteristics separately.

\paragraph{Datasets} We contrast complied benign inputs (from OR-Bench \citep{cui2024or}) with complied harmful inputs, ensuring both datasets exhibit similar compliance behavior but differ in harmfulness.
% \paragraph{Datasets} We selected some benign inputs from OR-Bench \citep{cui2024or} and combined them with compiled harmful inputs. This approach ensures that both datasets demonstrate similar compliance behavior while differing in their level of harmfulness. The sampled benign data will also be used for extracting compliance vectors.

\paragraph{Identifying HD} We apply the same difference-in-means to identify HD by calculating the mean activation $\boldsymbol{\mu}_{i,l}^{\text{\scriptsize c-benign}}$ for benign inputs from $D^{\text{\tiny compliance}}_{\text{\tiny benign}}$
\begin{align}
\boldsymbol{\mu}^{l}_{\text{\scriptsize c-benign}} = \frac{1}{|D^{\text{\tiny  compliance}}_{\text{\tiny  benign}}|} \sum_{x \in D^{\text{\tiny  compliance}}_{\text{\tiny  benign}}} \boldsymbol{h}^{l}(x)
\end{align}
Then HD is identified by:
\begin{align}
\label{equ:HD}
\boldsymbol{d}^{l}_{\text{HD}} &= \boldsymbol{\mu}^{l}_{\text{\scriptsize c-benign}} - \boldsymbol{\mu}^{l}_{\text{\scriptsize c-harmful}}
\end{align}

% \paragraph{Extracting compliance steering vector} 
% After applying the rejection vector, our goal is to reduce false rejections of benign inputs while minimizing any effects on harmful inputs. To accomplish this, we use the difference-in-means approach to create a specific false rejection vector for benign inputs. 
% \begin{align}
% \boldsymbol{\mu}^{l}_{\text{\scriptsize r-benign}} &= \frac{1}{|D^{\text{\tiny  rejection}}_{\text{\tiny  benign}}|} \sum\nolimits_{x \in D^{\text{\tiny  rejection}}_{\text{\tiny  benign}}} \boldsymbol{h}^{l}(x)\\
% \boldsymbol{d}^{l}_{\text{FRD}} &= \boldsymbol{\mu}^{l}_{\text{\scriptsize r-benign}} - \boldsymbol{\mu}^{l}_{\text{\scriptsize c-benign}}
% \end{align}
% where $\boldsymbol{d}^{l}_{\text{FRD}}$ represents the extracted false rejection vector, which we project onto RD and then take the negative to obtain the final compliance vector.
% \begin{equation}
% \label{equ:CV}
% \boldsymbol{v}_{\text{HD}} = -\boldsymbol{d}^{l}_{\text{FRD}} \boldsymbol{d}^{l}_{\text{RD}}{}^\top \boldsymbol{d}^{l}_{\text{HD}}
% \end{equation}

\paragraph{Extracting compliance steering vector} 
In fact, HD represents the backbone’s compliance behavior---identified by benign and harmful inputs that are both complied by the model---We can extract the compliance steering vector along HD to resist the influence of $\boldsymbol{v}^l_{\text{RD}}$, thereby mitigating the false rejection on benign inputs.

More specifically, we take the projection of $\boldsymbol{d}^l_{\text{HD}}$ along $\boldsymbol{d}^l_{\text{HD}}$ as the compliance steering vector, which assists in offsetting the rejection vector on benign inputs, thereby enhancing utility:
\begin{equation}
\label{equ:CV}
\boldsymbol{v}_{\text{HD}} = \boldsymbol{d}^{l}_{\text{RD}} \boldsymbol{d}^{l}_{\text{RD}}{}^\top \boldsymbol{d}^{l}_{\text{HD}}
\end{equation}

\paragraph{Deriving the Harmfulness Law}
As shown in Figure \ref{fig:llama31_dis}, along the HD direction (x-axis), we also define the Harmful Compliance Center (red point) as the origin. The leftward direction represents less harmful (positive), while the rightward direction represents increased harmfulness (negative). Each input is projected onto the HD, yielding a coordinate $pos_{\text{HD}}$. On HD, we notice that jailbreak inputs generally have smaller $pos_{\text{HD}}$ values, whereas benign inputs, tend to have larger $pos_{\text{HD}}$ values, which can be summarized as the following Harmfulness Law.
\begin{mdframed}[backgroundcolor=mycolor_orange!50, hidealllines=true]
\textbf{Harmfulness Law}: Inputs that are positioned further in the positive direction along HD require a greater compliance steering coefficient $\lambda_c$ to encourage compliance.
\end{mdframed}

\paragraph{Fitting the Harmfulness Law} Similar to RD, $pos_{\text{HD}}$ can be obtained by:
\begin{equation}
\label{equ:dis_hd}
pos_{\text{HD}} =   (\boldsymbol{h}^{l}-\boldsymbol{\mu}^{l}_{\text{\scriptsize c-harmful}}) \cdot d_{\text{HD}}^{l}
\end{equation}

For benign inputs from OR-Bench that are falsely rejected, we apply compliance steering vectors at varying intensities. For each input, we record its original $pos_{\text{HD}}$ and determine the $\lambda_c$ value required for the model to accept it. We fit a logistic regression curve to these ($pos_{\text{HD}}$, $\lambda_c$) pairs.
\begin{equation}
\label{eq:linear_c}
\lambda_{c} = w_c \cdot pos_{\text{HD}} + b_c
\end{equation}
where $w_c$, $b_c$ are parameters of logistic regression. Additionally, we conduct a small-scale grid search around the fitted hyperparameters.

\subsection{Adaptive Activation Steering}
Given any input prompt $t'$, we first utilize Eq. (\ref{eq:linear_r}) and Eq. (\ref{eq:linear_c}) to compute the steering coefficients $\lambda_r$ and $\lambda_c$ based on the positions $pos_{\text{RD}}$ and $pos_{\text{HD}}$. We then substitute these coefficients into Eq. (\ref{eq:adasteer}) to perform adaptive steering on the model’s hidden states across all layers at each token position, ensuring controlled safety behavior.

\section{Experiments}

\subsection{Experimental Setup}

\paragraph{Backbone} We conduct experiments on three aligned LLMs: LLaMA-3.1-8B-Instruct \citep{dubey2024llama}, Qwen2.5-7B-Instruct \citep{yang2024qwen2} and Gemma-2-9B-it \citep{team2024gemma} to evaluate the effectiveness of our approach.

\paragraph{Benchmark} We test our approach against several state-of-the-art jailbreak attack methods, including role-playing attacks, \textbf{AIM}%\footnote{https://jailbreakchat-hko42cs2r-alexalbertt-s-team.vercel.app/prompt/4f37a029-9dff-4862-b323-c96a5504de5d}
, gradient- or genetic algorithm-based prompt optimization techniques: \textbf{AutoDAN} \citep{liu2024autodan} and \textbf{GCG} \citep{zou2023universal}, and attacks that encrypt malicious queries using methods such as code, Base64 encoding, ciphering, LaTeX, and low-resource languages: \textbf{Jailbroken} \citep{wei2024jailbroken}, \textbf{Cipher} \citep{yuan2024cipher}, \textbf{ReNeLLM} \citep{DBLP:journals/corr/abs-2311-08268}, and \textbf{MultiLingual} \citep{deng2024multilingual}.
To assess utility, we employ over-safety test suites such as \textbf{XSTest} \citep{rottger-etal-2024-xstest} and \textbf{OKTest} \citep{shi-etal-2024-navigating}, along with the general instruction-following benchmark \textbf{AlpacaEval} \citep{dubois2024length}. Please refer to Appendix \ref{app:jailbreak} for details.

\paragraph{Metrics} For safety evaluation, we use the \textit{Defense Success Rate (DSR)}, which is computed using GPT-4o. For assessments on XSTest and OKTest, we follow \citet{rottger-etal-2024-xstest} and employ GPT-4o to measure the \textit{Compliance Rate (CR)}, representing the proportion of fully compliant responses. Additionally, we evaluate the general utility on AlpacaEval using the \textit{Win Rate}, which compares the quality of generated responses against the original model. A higher win rate indicates better preservation of the original model’s capabilities.

\begin{table*}[h]
\scriptsize
\centering
\begin{tabular}{l | c c c c c c c c | c | c}
\toprule
\textbf{}        & \multicolumn{8}{c|}{\textbf{Jailbreak Attack}} & \multicolumn{1}{c|}{\textbf{Over-Safety}} & \multicolumn{1}{c}{\textbf{Utility}} \\
% \midrule
\textbf{}        & \multicolumn{8}{c|}{\textbf{DSR}$\uparrow$} & \multicolumn{1}{c|}{\textbf{CR}$\uparrow$} & \multicolumn{1}{c}{\textbf{Win Rate}$\uparrow$} \\
% \midrule
\cmidrule(lr){2-9}
\cmidrule(lr){10-10}
\cmidrule(lr){11-11}
% &  \cellcolor{mycolor_red1}{AIM} &   \cellcolor{mycolor_red2}{AIM} &   \cellcolor{mycolor_red3}{AIM} &    GCG &   Jailbroken &    Multilingual &  ReNeLLM &\textbf{AVG.} &\textbf{AVG.} & AlpacaEval  \\
&   AIM &   AutoDAN &   Cipher &    GCG &   Jailbroken &    Multilingual &  ReNeLLM &\textbf{AVG.} &\textbf{AVG.} & AlpacaEval  \\
\midrule
LLaMA-3.1        &57  & 30  & 0   & 60  & 61  & 22  & 37  & 38.14 & 94.40 & 50.00 \\
\cmidrule(lr){1-1}
\cmidrule(lr){2-9}
\cmidrule(lr){10-10}
\cmidrule(lr){11-11}
ROSE        &\textbf{100} & 83  & 51  & \textbf{94}  & \textbf{85}  & 61  & 85  & 79.86 &90.47 &2.81 \\
Self-CD        &94  & 67  & 5   & 66  & 67  & 43  & 43  & 55.00  &93.74 &2.27 \\
Jailbreak Antidote        &92  & \textbf{100} & 61  & \textbf{94}  & 79  & 44  & 66  & 76.57 & 91.44 & 45.93  \\
Surgical        &\textbf{100} & 75  & 10  & 88  & 84  & 82  & \textbf{91}  & 75.71 &82.37 &47.29 \\
InferAligner        &85  & 90  & 0   & 92  & 77  & 82  & 77  & 71.86 &80.47&47.19 \\
CAST        &\textbf{100} & \textbf{100} & 0   & 66  & 76  & 46  & 56  & 63.43 &95.00 &37.76 \\
\midrule
% \textbf{AdaSteer (Ours)} &100 & 100 & 82  & 94  & 86  & 96  & 86  & 92.00 & 95.20 & 50.10 \\
\rowcolor{gray!20} \textbf{AdaSteer (Ours)} &\textbf{100} & \textbf{100} & \textbf{82}  & 90  & \textbf{85}  & \textbf{100}  & 86  & \textbf{91.86} & \textbf{97.87} & \textbf{50.01} \\
\midrule
\midrule
Qwen2.5        & 92  & 47  & 0   & 88  & 46  & 14  & 3   & 41.43 &95.00 &50.00 \\
\cmidrule(lr){1-1}
\cmidrule(lr){2-9}
\cmidrule(lr){10-10}
\cmidrule(lr){11-11}
ROSE        &        99  & 52  & 8   & 86  & 58  & 12  & 0   & 45.00 & \textbf{97.00} & 1.03 \\
 Self-CD        &       69  & 50  & 2   & 82  & 54  & 6   & 0   & 37.57 &96.00 & 0.96 \\
Jailbreak Antidote        &        88  & 86  & 72  & \textbf{100} & 60  & 78  & 3   & 69.57 &93.17 & 42.86 \\
 Surgical        &       94  & 41  & 0   & 82  & 47  & 13  & 3   & 40.00 & 95.24 & \textbf{48.85 }\\
  InferAligner        &      \textbf{100} & \textbf{98}  & 0   & 98  & 60  & \textbf{94}  & 11  & 65.86 &93.40 & 48.43 \\
 CAST        &       80  & 73  & 0   & 68  & 63  & 9   & 1   & 42.00 &95.60 & 47.90 \\
\midrule

\rowcolor{gray!20} \textbf{AdaSteer (Ours)} & \textbf{100} & \textbf{98} & \textbf{88}  & 92 & \textbf{78}  & 90 & \textbf{96}  & \textbf{91.71} &91.10 &48.36  \\

\midrule
\midrule
Gemma-2        &         6   & 31  & 0   & 90  & 57  & 1   & 27  & 30.29 &86.27 &50.00 \\
\cmidrule(lr){1-1}
\cmidrule(lr){2-9}
\cmidrule(lr){10-10}
\cmidrule(lr){11-11}
ROSE        &        7   & 50  & 25  & \textbf{100} & 67  & 20  & \textbf{87}  & 50.86 & 81.74 &1.98  \\
 Self-CD        &        4   & 25  & 0   & 90  & 56  & 0   & 46  & 31.57 &85.24&1.75 \\
Jailbreak Antidote        &         6   & 47  & 0   & 98  & 61  & 1   & 78  & 41.57&83.34&47.33  \\
 Surgical        &          \textbf{99}  & \textbf{100} & 14  & 98  & 68  & \textbf{96}  & 78  & 79.00 &90.57 &38.98 \\
InferAligner        &        31  & \textbf{100} & 24  & \textbf{100} & 85  & 93  & 62  & 70.71 &74.44 &48.48 \\
  CAST        &         8   & 35  & 0   & 94  & 65  & 4   & 33  & 34.14 &81.94 &\textbf{50.32}  \\
\midrule
\rowcolor{gray!20} \textbf{AdaSteer (Ours)} &         91  & 95  & \textbf{75}  & 86  & \textbf{86}  & 86  & 82  & \textbf{85.86} &\textbf{92.80} &48.28  \\
\bottomrule
\end{tabular}
\caption{The overall results of the three backbones (LLaMA-3.1-8B-Instruct, Qwen2.5-7B-Instruct, and Gemma-2-9B-it) on the benchmarks of jailbreak defense, over-safety, and model utility. The evaluation metric for jailbreak defense is the Defense Success Rate (DSR) for each attack method, the evaluation criterion for over-safety is the Compliance Rate (CR), and the utility is measured by the win rate compared to the original model.}
\label{tab:mainresults}
\end{table*}

\paragraph{Baselines and Comparison Methods}
We evaluate AdaSteer against the following training-free defense baselines, including Decoding-based Methods: (1) \textbf{ROSE} \citep{zhong2024rose}, (2) \textbf{Self-CD} \citep{shi2024navigating}
and Steering-based Methods: (3) \textbf{Jailbreak Antidote} \citep{shen2025jailbreak}, (4) \textbf{Surgical} \citep{wang2025surgical}, (5) \textbf{InferAligner} \citep{wang2024inferalignerinferencetimealignmentharmlessness}, (6) \textbf{CAST} \citep{lee2025programming}.
Please refer to Appendix \ref{app:baseline} for the detailed description.

\paragraph{Implementation Details}
We conduct experiments with PyTorch \citep{paszke2019pytorch} on a single NVIDIA Tesla A100 GPU. We set \texttt{do\_sample} to \texttt{False} for generation, which means using greedy decoding. Additional implementation details are provided in Appendix \ref{app:implementation}.

\begin{table*}
\scriptsize
\centering
\resizebox{\linewidth}{!}{
\begin{tabular}{l c | c c c c c c c | c c | c}
\toprule
\textbf{}   &     & \multicolumn{7}{c|}{\textbf{Jailbreak Attack}} & \multicolumn{2}{c|}{\textbf{Over-Safety}} & \multicolumn{1}{c}{\textbf{Utility}} \\
\cmidrule(lr){3-9}
\cmidrule(lr){10-11}
\cmidrule(lr){12-12}
& &  AIM &   AutoDAN &   Cipher &  GCG &   Jailbroken &    Multilingual &  ReNeLLM &XSTest & OKTest & AlpacaEval  \\
\midrule
% llama31
\multirow{2}{*}{$d_{\textbf{RD}}$} &  $pos_{\text{RD}}$ & -71.77 & -74.84 & -72.16 & -26.36 & -63.80 & -68.85 & -65.07 & -40.65 & -45.62  & -50.96\\

 & $\lambda_{r}$ & \cellcolor{mycolor_red1}-0.21 & \cellcolor{mycolor_red1}0.22 & \cellcolor{mycolor_red1}0.20 & \cellcolor{mycolor_red3}0.08 & \cellcolor{mycolor_red3}0.14 & \cellcolor{mycolor_red3}0.17 & \cellcolor{mycolor_red3}0.13 & \cellcolor{mycolor_red3}0.08 & \cellcolor{mycolor_red3}0.08 & \cellcolor{mycolor_red3}0.09 \\
 \midrule
 \multirow{2}{*}{$d_{\textbf{HD}}$} & $pos_{\text{HD}}$ & -17.51 &  -17.36  & -12.78 & -17.01 & -15.36 & -14.74 & -25.55 & 18.36 & 15.04 & 5.98   \\
  & $\lambda_{c}$ & 0.02 &  0.03 & \cellcolor{mycolor_green3}0.10 & 0.01 & 0.05 & 0.07 & -0.11 & \cellcolor{mycolor_green1}0.32 & \cellcolor{mycolor_green1}0.30 & \cellcolor{mycolor_green1}0.22 \\
\bottomrule
\end{tabular}
}
\caption{Results of the average positions and steering strength for complied inputs from different jailbreak methods and benign inputs on LLaMA-3.1.}
\label{tab:strength}
\vspace{-3mm}
\end{table*}

\subsection{Overall Results}

Table \ref{tab:mainresults} demonstrates the performance comparison of \textbf{AdaSteer} and baselines based on LLaMA-3.1-8B-Instruct, Qwen2.5-7B-Instruct and Gemma-2-9B-it. For the results of over-safety on each dataset, please refer to the Appendix \ref{app:over-safety}. 

AdaSteer significantly outperforms all baseline methods in jailbreak defense across various attack strategies, achieving near-complete resistance (DSR = 100) in most cases. This demonstrates the effectiveness of dynamically adjusting steering strength based on the characteristics of different jailbreak methods. In contrast, existing methods, including the most advanced Jailbreak Antidote and Surgical, show inconsistent performance across attack types, highlighting their vulnerability to certain adversarial techniques. Further, we adjust various hyperparameters for these two methods and identify a trade-off between safety, over-safety, and utility. By contrast, AdaSteer remains unaffected, underscoring our approach's superiority. Please refer to Appendix \ref{app:further_analysis} for detailed reuslts and analysis. The results validate our claim that a fixed steering struggles to generalize against diverse jailbreak attacks, while AdaSteer's adaptive mechanism ensures robust and comprehensive defense. 
% To better evaluate AdaSteer under adaptive attacks like AutoDAN and GCG, we apply them online to the protected model, which still shows strong defense. Please see Appendix \ref{app:online} for details.

Regarding benign inputs, AdaSteer maintains performance close to the original model, as reflected in its high utility win rate and strong compliance retention. This confirms its ability to distinguish between jailbreak and benign inputs, preserving model utility without over-enforcing refusals. Notably, while CAST applies conditional steering, its approach only differentiates between vanilla harmful prompts and benign queries, failing to effectively address jailbreak inputs due to their adversarial nature mimicking benign behavior. This limitation underscores the necessity of introducing Harmfulness Direction (HD) to separate jailbreak and benign inputs more effectively, further justifying our design choice in AdaSteer.

\subsection{Analysis of Adaptive Steering}
To directly demonstrate how AdaSteer operates, Table \ref{tab:strength} quantifies average $pos_{\text{RD}}$ and  $pos_{\text{HD}}$ for benign (AlpacaEval) and different types of jailbreak inputs on LLaMA-3.1, alongside the corresponding $\lambda_r$ and $\lambda_c$ computed by AdaSteer. The results indicate that: On $d_{\text{RD}}$, AdaSteer strongly rejects jailbreak inputs while minimizing rejection for benign queries. On $d_{\text{HD}}$, benign inputs receive a higher  $\lambda_c$, counteracting the rejection effect, while jailbreak inputs remain largely unaffected. Results for Qwen2.5 and Gemma-2 are in Appendix \ref{app:adaptive}.

\begin{table}
\centering
\small
\resizebox{\linewidth}{!}{
\begin{tabular}{l |c c c }
\toprule
\textbf{LLaMA-3.1}        & Jailbreak$\uparrow$ & Over-Safety$\uparrow$ & Utility$\uparrow$ \\
\midrule
\rowcolor{gray!20} \textbf{AdaSteer} &91.86 &97.87 &50.01 \\
\midrule
w/o $\boldsymbol{v}_{\text{RD}}$ & 39.57 &\textbf{98.54} &\textbf{50.70} \\
w/o $\boldsymbol{v}_{\text{HD}}$ & 91.57 & 74.37 & 45.72  \\
w/ reverse $\boldsymbol{v}_{\text{RD}}$ &\textbf{92.14} &95.20 &47.02 \\
\midrule
\midrule
\textbf{Qwen2.5}        & Jailbreak$\uparrow$  & Over-Safety$\uparrow$ & Utility$\uparrow$ \\
\midrule
\rowcolor{gray!20} \textbf{AdaSteer} &91.71 &91.10 &48.36 \\
\midrule
w/o $\boldsymbol{v}_{\text{RD}}$ & 46.00 &\textbf{96.54} & \textbf{48.82} \\
w/o $\boldsymbol{v}_{\text{HD}}$ & \textbf{92.86}   &79.60 &36.37 \\
w/ reverse $\boldsymbol{v}_{\text{RD}}$ &87.43 &90.54 &48.05 \\
\midrule
\midrule
\textbf{Gemma-2}        & Jailbreak$\uparrow$  & Over-Safety$\uparrow$ & Utility$\uparrow$ \\
\midrule
\rowcolor{gray!20} \textbf{AdaSteer} &85.86 &92.80 &48.28 \\
\midrule
w/o $\boldsymbol{v}_{\text{RD}}$ & 56.57 &88.67 & \textbf{49.99} \\
w/o $\boldsymbol{v}_{\text{HD}}$ & \textbf{92.14} & 90.17  & 33.08 \\
w/ reverse $\boldsymbol{v}_{\text{RD}}$ &91.43 &\textbf{96.60} &46.00 \\
\bottomrule
\end{tabular}
}
\caption{Ablation study on the effectiveness of steering vectors in our AdaSteer.}
\label{tab:ablation}
\vspace{-4mm}
\end{table}

\subsection{Steering Vector Analysis}

Tabel \ref{tab:ablation} presents the results of the ablation study evaluating the impact of different steering vectors in AdaSteer across three backbones. We compare the full AdaSteer method with three ablated versions: (1) w/o $\boldsymbol{v}_{\text{RD}}$, which removes rejection steering, (2) w/o $\boldsymbol{v}_{\text{HD}}$, which removes compliance steering, and (3) w/ reverse $\boldsymbol{v}_{\text{RD}}$, which replaces $\boldsymbol{v}_{\text{HD}}$ with the inverted $\boldsymbol{v}_{\text{RD}}$.

The results show that removing $\boldsymbol{v}_{\text{RD}}$ lowers jailbreak resistance, confirming its role in reinforcing rejection behavior. Conversely, removing $\boldsymbol{v}_{\text{HD}}$ significantly degrades utility, indicating that compliance steering is crucial for reducing false rejections. The reverse $\boldsymbol{v}_{\text{RD}}$  setting achieves comparable jailbreak defense but sacrifices utility, demonstrating that simply inverting the rejection vector is suboptimal for distinguishing benign inputs. These findings validate the necessity of steering along both rejection and harmfulness direction for achieving robust and adaptive jailbreak defense.

\begin{figure}
  \centering
  \includegraphics[width=\linewidth]{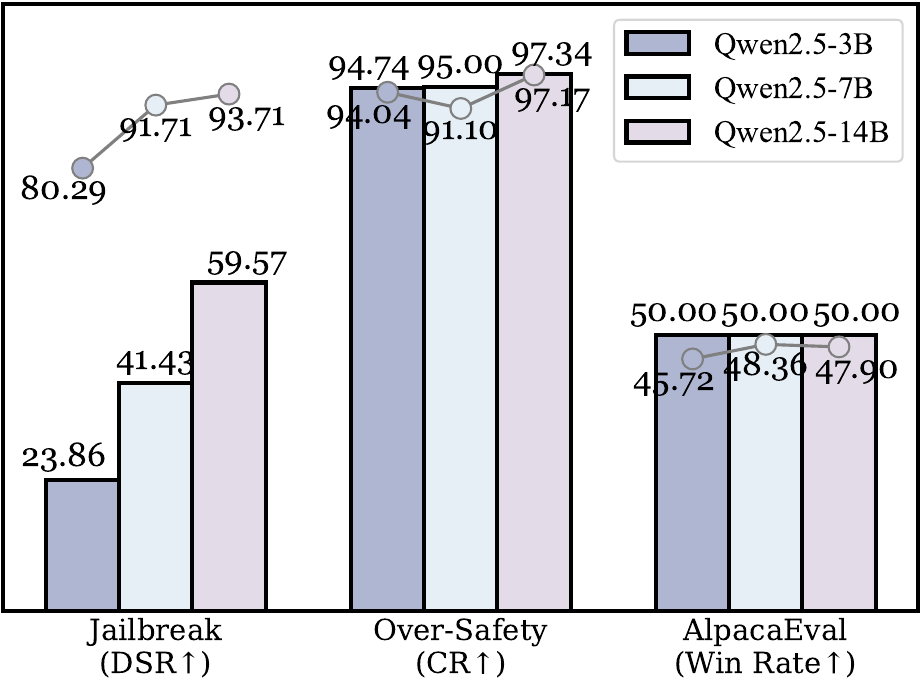} 
  \caption{The results of AdaSteer across different sizes of Qwen2.5. The values above the bars represent the original model’s performance, while the values below the line indicate that after applying AdaSteer.}
  \label{fig:qwen25_size}
\vspace{-3mm}
\end{figure}

\subsection{The Impact of Model Size}

To evaluate the scalability of AdaSteer, we assess it across three different sizes of Qwen2.5 models ranging from 3B to 14B, as shown in Figure \ref{fig:qwen25_size}. The results demonstrate that AdaSteer significantly enhances jailbreak defense across all model sizes while maintaining performance on benign inputs, highlighting its adaptability to different model capacities. This consistency across scales underscores AdaSteer’s robustness as a generalizable safety enhancement method. Moreover, the results reveal that even smaller models, which are typically more vulnerable to jailbreak attacks, can leverage AdaSteer to achieve significant improvement on adaptive jailbreak defense. This suggests that adaptive jailbreak defense is not exclusive to large-scale models---smaller models, when equipped with our AdaSteer, can also exhibit strong adversarial robustness. Please refer to Appendix \ref{app:vector} for the detailed results on each jailbreak type.

\begin{figure}
  \centering
  \includegraphics[width=\linewidth]{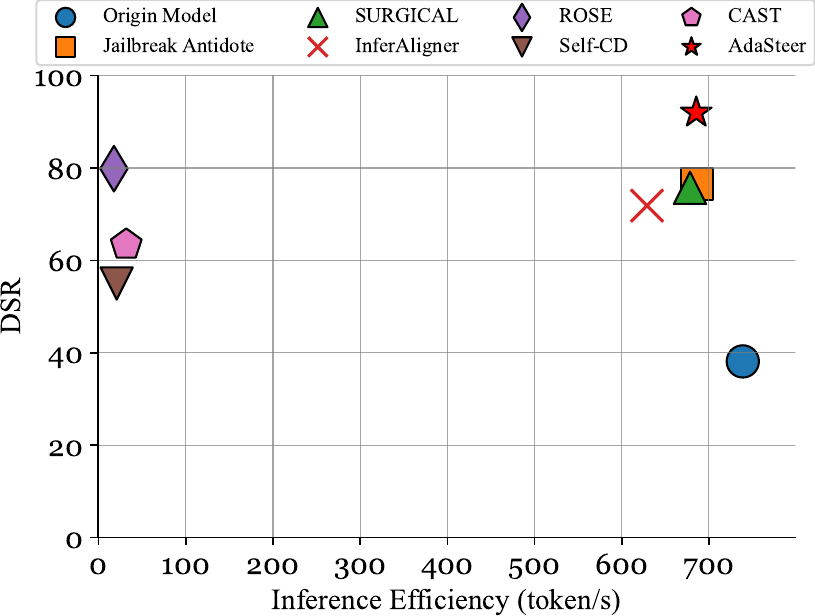}
  \caption{Trade-off between inference efficiency and jailbreak defense success rate (DSR).}
  \label{fig:llama31_time}
\vspace{-4mm}
\end{figure}

\subsection{Inference Efficiency Analysis}

To evaluate the efficiency of different jailbreak defense methods, we compare their tokens per second (token/s) relative to the original model.
% while maintaining a uniform batch size of 64 for methods that support batch inference. 
We conduct our experiments on a single NVIDIA Tesla A100 GPU. For methods that support batch inference, we set the batch size to 64.
% For methods that do not support batching, we use the ratio of their total inference time to that of the original model as an efficiency indicator. 
The trade-off between inference efficiency and jailbreak defense success rate (DSR) is visualized in Figure \ref{fig:llama31_time}. AdaSteer is positioned in the upper-right region of the plot, demonstrating that it achieves a strong balance between safety and efficiency. 
% Unlike other high-performing defenses that introduce significant computational overhead, AdaSteer maintains high DSR without excessive inference cost.
Unlike other high-performing defenses that introduce significant computational overhead, AdaSteer maintains high DSR without excessive inference cost, preserving a runtime speed close to that of the original model.
This highlights its practicality as a scalable and efficient solution for enhancing model security in real-world deployments.

\section{Related Works}

\paragraph{Jailbreak Attack}
Recent studies have exposed a significant threat termed jailbreak attack, where adversarial prompts are designed to bypass safety mechanisms and induce models to generate harmful content. Existing jailbreak methods can be classified into three types \citep{zhou2024easyjailbreak}: (1) Human Design \cite{li2023multi,li2023deepinception,shayegani2023jailbreak,wei2023jailbreak}, which encompasses jailbreak prompts crafted manually, leveraging human creativity to bypass safeguards (2) Long-tail Encoding \cite{yuan2023gpt,deng2024multilingual,lv2024codechameleon}, which leverages the limited cross-task generalization ability of LLMs to unseen data during safety alignment, and (3) Prompt Optimization \cite{zou2023universal,liu2023autodan,yu2023gptfuzzer,chao2023jailbreaking,ding2023wolf,mu2024stealthy} aims at automatically designing jailbreak prompt to induce harmful content. These diverse attacks highlight the urgent need for robust and flexible defenses to maintain LLM safety.

\paragraph{Jailbreak Defense}
Safety post-training is a widely used approach for enhancing LLMs' resistance to jailbreak attacks. Some methods strengthen the model’s refusal behavior by further fine-tuning on safety data \citep{xu2024safedecoding,zhao2024towards} or applying preference optimization \citep{bai2022training,ouyang2022training,rafailov2023direct}. Others employ machine unlearning techniques \citep{yao2023large,liu2024towards,zhang2024safe} to erase harmful knowledge from the model. However, these approaches often come with substantial computational costs and are highly sensitive to variations in training data, resulting in inconsistent performance.

\paragraph{Activation Steering} Steering representation within LLMs has garnered increasing attention due to its transparency and lightweight properties \citep{zou2023representation}. This technique is grounded in the theoretical premise that LLMs encode features or concepts as linear directions in activation space \citep{mikolov2013linguistic,park2024linear}. Exist works mainly adopt static steering with a fixed coefficient exerted on the extracted refusal vectors for jailbreak defense \citep{zheng2024prompt,qian2024hsf,stickland2024steering,li2025revisiting,shen2025jailbreak}. Although few works explore more fine-grained steering control, they are still narrowed within vanilla harmful prompt scenario \citep{bhattacharjee2024towards,wang2024adaptive,lee2025programming}, leaving the more challenging jailbreak attacks under-explored.

AdaSteer stands out by enabling dynamic and input-dependent control over jailbreak defenses, effectively enhancing safety while preserving utility.

\section{Conclusion}
In this work, we propose AdaSteer, a dual-direction adaptive activation steering method that enhances jailbreak defense in LLMs while maintaining their utility. By identifying two key properties---Rejection Law and Harmfulness Law---we show that jailbreak inputs exhibit distinct behaviors in activation space, allowing for dynamic, input-aware steering along the Rejection and Harmfulness Direction.
Extensive experiments on LLaMA-3.1, Gemma-2, and Qwen2.5 confirm that AdaSteer outperforms baselines across diverse jailbreak strategies, demonstrating its effectiveness and scalability.

\section*{Limitations}
Despite the effectiveness of AdaSteer, our study has certain limitations that warrant further exploration.

First, due to computational constraints, our experiments are conducted on mid-sized LLMs (e.g., LLaMA-3.1-8B, Gemma-2-9B, and Qwen2.5-7B). While our results demonstrate the scalability of AdaSteer across different model sizes, its performance on larger-scale models (e.g., 30B+ parameters) remains unverified. Future work should investigate whether AdaSteer maintains its efficiency and adaptability in frontier LLMs.

Second, our method relies on linear activation steering, assuming that model behaviors can be effectively controlled via low-dimensional vector manipulations. While this has shown strong empirical results, future research could explore nonlinear adaptations or layer-wise adjustments to further refine AdaSteer’s adaptability.

Despite these limitations, our findings demonstrate the practicality, efficiency, and robustness of AdaSteer, paving the way for scalable and interpretable jailbreak defenses in LLMs.

\section*{Ethical Considerations}
Our work is conducted solely for research purposes and aims to enhance the security and robustness of LLMs against adversarial jailbreak attacks. AdaSteer is designed to improve model alignment with human values by providing an adaptive, interpretable, and training-free defense mechanism. Our study does not intend to create or facilitate new jailbreak techniques but rather to understand and mitigate existing vulnerabilities in LLMs.

Furthermore, our research focuses on interpreting the internal safety mechanisms of LLMs, contributing to the broader goal of responsible AI development. The datasets used in our experiments are publicly available and widely adopted in the field. We strictly adhere to ethical guidelines, ensuring that our methodology does not promote or reinforce harmful behaviors.

While AdaSteer improves jailbreak defense, no security measure is absolute. We encourage continued collaborative research on evolving safety threats and emphasize the importance of transparent, ethical AI deployment to safeguard LLM usage in real-world applications.

\section*{Acknowledgments}
We thank the anonymous reviewers for their comments and suggestions. This work was supported by the New Generation Artificial Intelligence-National Science and Technology Major Project 2023ZD0121100, the National Natural Science Foundation of China (NSFC) via grant 62441614 and 62176078, the Fundamental Research Funds for the Central Universities, and the Singapore Ministry of Education (MOE) Academic Research Fund (AcRF) Tier 1 grant (No. MSS24C012).


\bibliography{custom}

\begin{thebibliography}{69}
\providecommand{\natexlab}[1]{#1}

\bibitem[{Anthropic(2025)}]{anthropic2025blog}
Anthropic. 2025.
\newblock \href {https://alignment.anthropic.com/2025/recommended-directions/#h.a2hnl4sly5t5} {Recommendations for technical ai safety research directions}.
\newblock \emph{Anthropic's Alignment Science Blog}.

\bibitem[{Arditi et~al.(2024)Arditi, Obeso, Syed, Paleka, Panickssery, Gurnee, and Nanda}]{arditi2024refusal}
Andy Arditi, Oscar Obeso, Aaquib Syed, Daniel Paleka, Nina Panickssery, Wes Gurnee, and Neel Nanda. 2024.
\newblock Refusal in language models is mediated by a single direction.
\newblock \emph{arXiv preprint arXiv:2406.11717}.

\bibitem[{Askell et~al.(2021)Askell, Bai, Chen, Drain, Ganguli, Henighan, Jones, Joseph, Mann, DasSarma et~al.}]{askell2021general}
Amanda Askell, Yuntao Bai, Anna Chen, Dawn Drain, Deep Ganguli, Tom Henighan, Andy Jones, Nicholas Joseph, Ben Mann, Nova DasSarma, et~al. 2021.
\newblock A general language assistant as a laboratory for alignment.
\newblock \emph{arXiv preprint arXiv:2112.00861}.

\bibitem[{Bai et~al.(2022{\natexlab{a}})Bai, Jones, Ndousse, Askell, Chen, DasSarma, Drain, Fort, Ganguli, Henighan et~al.}]{bai2022training}
Yuntao Bai, Andy Jones, Kamal Ndousse, Amanda Askell, Anna Chen, Nova DasSarma, Dawn Drain, Stanislav Fort, Deep Ganguli, Tom Henighan, et~al. 2022{\natexlab{a}}.
\newblock Training a helpful and harmless assistant with reinforcement learning from human feedback.
\newblock \emph{arXiv preprint arXiv:2204.05862}.

\bibitem[{Bai et~al.(2022{\natexlab{b}})Bai, Kadavath, Kundu, Askell, Kernion, Jones, Chen, Goldie, Mirhoseini, McKinnon et~al.}]{bai2022constitutional}
Yuntao Bai, Saurav Kadavath, Sandipan Kundu, Amanda Askell, Jackson Kernion, Andy Jones, Anna Chen, Anna Goldie, Azalia Mirhoseini, Cameron McKinnon, et~al. 2022{\natexlab{b}}.
\newblock Constitutional ai: Harmlessness from ai feedback.
\newblock \emph{arXiv preprint arXiv:2212.08073}.

\bibitem[{Belrose(2023)}]{belrose2023diff}
Nora Belrose. 2023.
\newblock Diff-in-means concept editing is worst-case optimal: Explaining a result by sam marks and max tegmark, 2023.
\newblock \emph{URL https://blog. eleuther. ai/diff-in-means}.

\bibitem[{Bhattacharjee et~al.(2024)Bhattacharjee, Ghosh, Rebedea, and Parisien}]{bhattacharjee2024towards}
Amrita Bhattacharjee, Shaona Ghosh, Traian Rebedea, and Christopher Parisien. 2024.
\newblock Towards inference-time category-wise safety steering for large language models.
\newblock In \emph{Neurips Safe Generative AI Workshop 2024}.

\bibitem[{Carlini et~al.(2024)Carlini, Nasr, Choquette-Choo, Jagielski, Gao, Koh, Ippolito, Tramer, and Schmidt}]{carlini2024aligned}
Nicholas Carlini, Milad Nasr, Christopher~A Choquette-Choo, Matthew Jagielski, Irena Gao, Pang Wei~W Koh, Daphne Ippolito, Florian Tramer, and Ludwig Schmidt. 2024.
\newblock Are aligned neural networks adversarially aligned?
\newblock \emph{Advances in Neural Information Processing Systems}, 36.

\bibitem[{Chao et~al.(2024)Chao, Debenedetti, Robey, Andriushchenko, Croce, Sehwag, Dobriban, Flammarion, Pappas, Tramer et~al.}]{chao2024jailbreakbench}
Patrick Chao, Edoardo Debenedetti, Alexander Robey, Maksym Andriushchenko, Francesco Croce, Vikash Sehwag, Edgar Dobriban, Nicolas Flammarion, George~J Pappas, Florian Tramer, et~al. 2024.
\newblock Jailbreakbench: An open robustness benchmark for jailbreaking large language models.
\newblock \emph{arXiv preprint arXiv:2404.01318}.

\bibitem[{Chao et~al.(2023)Chao, Robey, Dobriban, Hassani, Pappas, and Wong}]{chao2023jailbreaking}
Patrick Chao, Alexander Robey, Edgar Dobriban, Hamed Hassani, George~J Pappas, and Eric Wong. 2023.
\newblock Jailbreaking black box large language models in twenty queries.
\newblock In \emph{R0-FoMo: Robustness of Few-shot and Zero-shot Learning in Large Foundation Models}.

\bibitem[{Cui et~al.(2024)Cui, Chiang, Stoica, and Hsieh}]{cui2024or}
Justin Cui, Wei-Lin Chiang, Ion Stoica, and Cho-Jui Hsieh. 2024.
\newblock Or-bench: An over-refusal benchmark for large language models.
\newblock \emph{arXiv preprint arXiv:2405.20947}.

\bibitem[{Deng et~al.(2023)Deng, Liu, Li, Wang, Zhang, Li, Wang, Zhang, and Liu}]{deng2023jailbreaker}
Gelei Deng, Yi~Liu, Yuekang Li, Kailong Wang, Ying Zhang, Zefeng Li, Haoyu Wang, Tianwei Zhang, and Yang Liu. 2023.
\newblock Jailbreaker: Automated jailbreak across multiple large language model chatbots.
\newblock \emph{arXiv preprint arXiv:2307.08715}.

\bibitem[{Deng et~al.(2024)Deng, Zhang, Pan, and Bing}]{deng2024multilingual}
Yue Deng, Wenxuan Zhang, Sinno~Jialin Pan, and Lidong Bing. 2024.
\newblock Multilingual jailbreak challenges in large language models.
\newblock In \emph{The Twelfth International Conference on Learning Representations}.

\bibitem[{Ding et~al.(2023{\natexlab{a}})Ding, Kuang, Ma, Cao, Xian, Chen, and Huang}]{DBLP:journals/corr/abs-2311-08268}
Peng Ding, Jun Kuang, Dan Ma, Xuezhi Cao, Yunsen Xian, Jiajun Chen, and Shujian Huang. 2023{\natexlab{a}}.
\newblock A wolf in sheep's clothing: Generalized nested jailbreak prompts can fool large language models easily.
\newblock \emph{CoRR}, abs/2311.08268.

\bibitem[{Ding et~al.(2023{\natexlab{b}})Ding, Kuang, Ma, Cao, Xian, Chen, and Huang}]{ding2023wolf}
Peng Ding, Jun Kuang, Dan Ma, Xuezhi Cao, Yunsen Xian, Jiajun Chen, and Shujian Huang. 2023{\natexlab{b}}.
\newblock A wolf in sheep's clothing: Generalized nested jailbreak prompts can fool large language models easily.
\newblock \emph{arXiv preprint arXiv:2311.08268}.

\bibitem[{Dubey et~al.(2024)Dubey, Jauhri, Pandey, Kadian, Al-Dahle, Letman, Mathur, Schelten, Yang, Fan et~al.}]{dubey2024llama}
Abhimanyu Dubey, Abhinav Jauhri, Abhinav Pandey, Abhishek Kadian, Ahmad Al-Dahle, Aiesha Letman, Akhil Mathur, Alan Schelten, Amy Yang, Angela Fan, et~al. 2024.
\newblock The llama 3 herd of models.
\newblock \emph{arXiv preprint arXiv:2407.21783}.

\bibitem[{Dubois et~al.(2024)Dubois, Galambosi, Liang, and Hashimoto}]{dubois2024length}
Yann Dubois, Bal{\'a}zs Galambosi, Percy Liang, and Tatsunori~B Hashimoto. 2024.
\newblock Length-controlled alpacaeval: A simple way to debias automatic evaluators.
\newblock \emph{arXiv preprint arXiv:2404.04475}.

\bibitem[{Huang et~al.(2024)Huang, Gupta, Xia, Li, and Chen}]{huang2024catastrophic}
Yangsibo Huang, Samyak Gupta, Mengzhou Xia, Kai Li, and Danqi Chen. 2024.
\newblock Catastrophic jailbreak of open-source {LLM}s via exploiting generation.
\newblock In \emph{The Twelfth International Conference on Learning Representations}.

\bibitem[{Jones et~al.(2023)Jones, Dragan, Raghunathan, and Steinhardt}]{jones2023automatically}
Erik Jones, Anca Dragan, Aditi Raghunathan, and Jacob Steinhardt. 2023.
\newblock Automatically auditing large language models via discrete optimization.
\newblock In \emph{International Conference on Machine Learning}, pages 15307--15329. PMLR.

\bibitem[{Lee et~al.(2025)Lee, Padhi, Ramamurthy, Miehling, Dognin, Nagireddy, and Dhurandhar}]{lee2025programming}
Bruce~W Lee, Inkit Padhi, Karthikeyan~Natesan Ramamurthy, Erik Miehling, Pierre Dognin, Manish Nagireddy, and Amit Dhurandhar. 2025.
\newblock \href {https://openreview.net/forum?id=Oi47wc10sm} {Programming refusal with conditional activation steering}.
\newblock In \emph{The Thirteenth International Conference on Learning Representations}.

\bibitem[{Leong et~al.(2024)Leong, Cheng, Xu, Wang, Wang, and Li}]{leong2024no}
Chak~Tou Leong, Yi~Cheng, Kaishuai Xu, Jian Wang, Hanlin Wang, and Wenjie Li. 2024.
\newblock No two devils alike: Unveiling distinct mechanisms of fine-tuning attacks.
\newblock \emph{arXiv preprint arXiv:2405.16229}.

\bibitem[{Li et~al.(2023{\natexlab{a}})Li, Guo, Fan, Xu, Huang, Meng, and Song}]{li2023multi}
Haoran Li, Dadi Guo, Wei Fan, Mingshi Xu, Jie Huang, Fanpu Meng, and Yangqiu Song. 2023{\natexlab{a}}.
\newblock Multi-step jailbreaking privacy attacks on chatgpt.
\newblock In \emph{Findings of the Association for Computational Linguistics: EMNLP 2023}, pages 4138--4153.

\bibitem[{Li et~al.(2025)Li, Wang, Liu, Wu, Dou, Lv, Wang, Zheng, and Huang}]{li2025revisiting}
Tianlong Li, Zhenghua Wang, Wenhao Liu, Muling Wu, Shihan Dou, Changze Lv, Xiaohua Wang, Xiaoqing Zheng, and Xuan-Jing Huang. 2025.
\newblock Revisiting jailbreaking for large language models: A representation engineering perspective.
\newblock In \emph{Proceedings of the 31st International Conference on Computational Linguistics}, pages 3158--3178.

\bibitem[{Li et~al.(2023{\natexlab{b}})Li, Zhou, Zhu, Yao, Liu, and Han}]{li2023deepinception}
Xuan Li, Zhanke Zhou, Jianing Zhu, Jiangchao Yao, Tongliang Liu, and Bo~Han. 2023{\natexlab{b}}.
\newblock Deepinception: Hypnotize large language model to be jailbreaker.
\newblock \emph{arXiv preprint arXiv:2311.03191}.

\bibitem[{Liu et~al.(2023)Liu, Xu, Chen, and Xiao}]{liu2023autodan}
Xiaogeng Liu, Nan Xu, Muhao Chen, and Chaowei Xiao. 2023.
\newblock Autodan: Generating stealthy jailbreak prompts on aligned large language models.
\newblock \emph{arXiv preprint arXiv:2310.04451}.

\bibitem[{Liu et~al.(2024{\natexlab{a}})Liu, Xu, Chen, and Xiao}]{liu2024autodan}
Xiaogeng Liu, Nan Xu, Muhao Chen, and Chaowei Xiao. 2024{\natexlab{a}}.
\newblock Auto{DAN}: Generating stealthy jailbreak prompts on aligned large language models.
\newblock In \emph{The Twelfth International Conference on Learning Representations}.

\bibitem[{Liu et~al.(2024{\natexlab{b}})Liu, Dou, Tan, Tian, and Jiang}]{liu2024towards}
Zheyuan Liu, Guangyao Dou, Zhaoxuan Tan, Yijun Tian, and Meng Jiang. 2024{\natexlab{b}}.
\newblock Towards safer large language models through machine unlearning.
\newblock \emph{arXiv preprint arXiv:2402.10058}.

\bibitem[{Lv et~al.(2024)Lv, Wang, Zhang, Huang, Dou, Ye, Gui, Zhang, and Huang}]{lv2024codechameleon}
Huijie Lv, Xiao Wang, Yuansen Zhang, Caishuang Huang, Shihan Dou, Junjie Ye, Tao Gui, Qi~Zhang, and Xuanjing Huang. 2024.
\newblock Codechameleon: Personalized encryption framework for jailbreaking large language models.
\newblock \emph{arXiv preprint arXiv:2402.16717}.

\bibitem[{Mazeika et~al.(2023)Mazeika, Hendrycks, Li, Xu, Hough, Zou, Rajabi, Yao, Wang, Tian et~al.}]{mazeika2023trojan}
Mantas Mazeika, Dan Hendrycks, Huichen Li, Xiaojun Xu, Sidney Hough, Andy Zou, Arezoo Rajabi, Qi~Yao, Zihao Wang, Jian Tian, et~al. 2023.
\newblock The trojan detection challenge.
\newblock In \emph{NeurIPS 2022 Competition Track}, pages 279--291. PMLR.

\bibitem[{Mazeika et~al.(2024)Mazeika, Phan, Yin, Zou, Wang, Mu, Sakhaee, Li, Basart, Li, Forsyth, and Hendrycks}]{mazeika2024harmbench}
Mantas Mazeika, Long Phan, Xuwang Yin, Andy Zou, Zifan Wang, Norman Mu, Elham Sakhaee, Nathaniel Li, Steven Basart, Bo~Li, David Forsyth, and Dan Hendrycks. 2024.
\newblock Harmbench: A standardized evaluation framework for automated red teaming and robust refusal.
\newblock In \emph{Forty-first International Conference on Machine Learning}.

\bibitem[{Mikolov et~al.(2013)Mikolov, Yih, and Zweig}]{mikolov2013linguistic}
Tom{\'a}{\v{s}} Mikolov, Wen-tau Yih, and Geoffrey Zweig. 2013.
\newblock Linguistic regularities in continuous space word representations.
\newblock In \emph{Proceedings of the 2013 conference of the north american chapter of the association for computational linguistics: Human language technologies}, pages 746--751.

\bibitem[{Mu et~al.(2024)Mu, He, Zhou, Feng, Xu, Qin, Shi, Liu, Han, Shi et~al.}]{mu2024stealthy}
Honglin Mu, Han He, Yuxin Zhou, Yunlong Feng, Yang Xu, Libo Qin, Xiaoming Shi, Zeming Liu, Xudong Han, Qi~Shi, et~al. 2024.
\newblock Stealthy jailbreak attacks on large language models via benign data mirroring.
\newblock \emph{arXiv preprint arXiv:2410.21083}.

\bibitem[{Ouyang et~al.(2022)Ouyang, Wu, Jiang, Almeida, Wainwright, Mishkin, Zhang, Agarwal, Slama, Ray et~al.}]{ouyang2022training}
Long Ouyang, Jeffrey Wu, Xu~Jiang, Diogo Almeida, Carroll Wainwright, Pamela Mishkin, Chong Zhang, Sandhini Agarwal, Katarina Slama, Alex Ray, et~al. 2022.
\newblock Training language models to follow instructions with human feedback.
\newblock \emph{Advances in neural information processing systems}, 35:27730--27744.

\bibitem[{Panickssery et~al.(2023)Panickssery, Gabrieli, Schulz, Tong, Hubinger, and Turner}]{panickssery2023steering}
Nina Panickssery, Nick Gabrieli, Julian Schulz, Meg Tong, Evan Hubinger, and Alexander~Matt Turner. 2023.
\newblock Steering llama 2 via contrastive activation addition.
\newblock \emph{arXiv preprint arXiv:2312.06681}.

\bibitem[{Park et~al.(2024)Park, Choe, and Veitch}]{park2024linear}
Kiho Park, Yo~Joong Choe, and Victor Veitch. 2024.
\newblock The linear representation hypothesis and the geometry of large language models.
\newblock In \emph{Forty-first International Conference on Machine Learning}.

\bibitem[{Paszke et~al.(2019)Paszke, Gross, Massa, Lerer, Bradbury, Chanan, Killeen, Lin, Gimelshein, Antiga et~al.}]{paszke2019pytorch}
Adam Paszke, Sam Gross, Francisco Massa, Adam Lerer, James Bradbury, Gregory Chanan, Trevor Killeen, Zeming Lin, Natalia Gimelshein, Luca Antiga, et~al. 2019.
\newblock Pytorch: An imperative style, high-performance deep learning library.
\newblock \emph{Advances in neural information processing systems}, 32.

\bibitem[{Qian et~al.(2024)Qian, Zhang, Sha, and Zheng}]{qian2024hsf}
Cheng Qian, Hainan Zhang, Lei Sha, and Zhiming Zheng. 2024.
\newblock Hsf: Defending against jailbreak attacks with hidden state filtering.
\newblock \emph{arXiv preprint arXiv:2409.03788}.

\bibitem[{Rafailov et~al.(2023)Rafailov, Sharma, Mitchell, Manning, Ermon, and Finn}]{rafailov2023direct}
Rafael Rafailov, Archit Sharma, Eric Mitchell, Christopher~D Manning, Stefano Ermon, and Chelsea Finn. 2023.
\newblock Direct preference optimization: Your language model is secretly a reward model.
\newblock \emph{Advances in Neural Information Processing Systems}, 36.

\bibitem[{R{\"o}ttger et~al.(2024)R{\"o}ttger, Kirk, Vidgen, Attanasio, Bianchi, and Hovy}]{rottger-etal-2024-xstest}
Paul R{\"o}ttger, Hannah Kirk, Bertie Vidgen, Giuseppe Attanasio, Federico Bianchi, and Dirk Hovy. 2024.
\newblock {XST}est: A test suite for identifying exaggerated safety behaviours in large language models.
\newblock In \emph{Proceedings of the 2024 Conference of the North American Chapter of the Association for Computational Linguistics: Human Language Technologies (Volume 1: Long Papers)}, pages 5377--5400.

\bibitem[{Shayegani et~al.(2023)Shayegani, Dong, and Abu-Ghazaleh}]{shayegani2023jailbreak}
Erfan Shayegani, Yue Dong, and Nael Abu-Ghazaleh. 2023.
\newblock Jailbreak in pieces: Compositional adversarial attacks on multi-modal language models.
\newblock In \emph{The Twelfth International Conference on Learning Representations}.

\bibitem[{Shen et~al.(2025)Shen, Zhao, Dong, He, and Zeng}]{shen2025jailbreak}
Guobin Shen, Dongcheng Zhao, Yiting Dong, Xiang He, and Yi~Zeng. 2025.
\newblock \href {https://openreview.net/forum?id=s20W12XTF8} {Jailbreak antidote: Runtime safety-utility balance via sparse representation adjustment in large language models}.
\newblock In \emph{The Thirteenth International Conference on Learning Representations}.

\bibitem[{Shi et~al.(2024{\natexlab{a}})Shi, Wang, Ge, Gao, Yang, Gui, Zhang, Huang, Zhao, and Lin}]{shi-etal-2024-navigating}
Chenyu Shi, Xiao Wang, Qiming Ge, Songyang Gao, Xianjun Yang, Tao Gui, Qi~Zhang, Xuanjing Huang, Xun Zhao, and Dahua Lin. 2024{\natexlab{a}}.
\newblock Navigating the {O}ver{K}ill in large language models.
\newblock In \emph{Proceedings of the 62nd Annual Meeting of the Association for Computational Linguistics (Volume 1: Long Papers)}, pages 4602--4614.

\bibitem[{Shi et~al.(2024{\natexlab{b}})Shi, Wang, Ge, Gao, Yang, Gui, Zhang, Huang, Zhao, and Lin}]{shi2024navigating}
Chenyu Shi, Xiao Wang, Qiming Ge, Songyang Gao, Xianjun Yang, Tao Gui, Qi~Zhang, Xuanjing Huang, Xun Zhao, and Dahua Lin. 2024{\natexlab{b}}.
\newblock Navigating the overkill in large language models.
\newblock \emph{arXiv preprint arXiv:2401.17633}.

\bibitem[{Stickland et~al.(2024)Stickland, Lyzhov, Pfau, Mahdi, and Bowman}]{stickland2024steering}
Asa~Cooper Stickland, Alexander Lyzhov, Jacob Pfau, Salsabila Mahdi, and Samuel~R Bowman. 2024.
\newblock Steering without side effects: Improving post-deployment control of language models.
\newblock \emph{arXiv preprint arXiv:2406.15518}.

\bibitem[{Team et~al.(2024)Team, Riviere, Pathak, Sessa, Hardin, Bhupatiraju, Hussenot, Mesnard, Shahriari, Ram{\'e} et~al.}]{team2024gemma}
Gemma Team, Morgane Riviere, Shreya Pathak, Pier~Giuseppe Sessa, Cassidy Hardin, Surya Bhupatiraju, L{\'e}onard Hussenot, Thomas Mesnard, Bobak Shahriari, Alexandre Ram{\'e}, et~al. 2024.
\newblock Gemma 2: Improving open language models at a practical size.
\newblock \emph{arXiv preprint arXiv:2408.00118}.

\bibitem[{Turner et~al.(2023)Turner, Thiergart, Leech, Udell, Vazquez, Mini, and MacDiarmid}]{turner2023activation}
Alexander~Matt Turner, Lisa Thiergart, Gavin Leech, David Udell, Juan~J Vazquez, Ulisse Mini, and Monte MacDiarmid. 2023.
\newblock Activation addition: Steering language models without optimization.
\newblock \emph{arXiv e-prints}, pages arXiv--2308.

\bibitem[{Wang et~al.(2024{\natexlab{a}})Wang, Mehrabi, Goyal, Gupta, Chang, and Galstyan}]{wang2024data}
Fei Wang, Ninareh Mehrabi, Palash Goyal, Rahul Gupta, Kai-Wei Chang, and Aram Galstyan. 2024{\natexlab{a}}.
\newblock Data advisor: Dynamic data curation for safety alignment of large language models.
\newblock In \emph{Proceedings of the 2024 Conference on Empirical Methods in Natural Language Processing}, pages 8089--8100.

\bibitem[{Wang et~al.(2024{\natexlab{b}})Wang, Zhang, Li, Tan, Wang, Ren, Jiang, and Qiu}]{wang2024inferalignerinferencetimealignmentharmlessness}
Pengyu Wang, Dong Zhang, Linyang Li, Chenkun Tan, Xinghao Wang, Ke~Ren, Botian Jiang, and Xipeng Qiu. 2024{\natexlab{b}}.
\newblock \href {https://arxiv.org/abs/2401.11206} {Inferaligner: Inference-time alignment for harmlessness through cross-model guidance}.
\newblock \emph{Preprint}, arXiv:2401.11206.

\bibitem[{Wang et~al.(2024{\natexlab{c}})Wang, Jiao, He, Chen, Zhu, Chu, Gao, Wang, and Ma}]{wang2024adaptive}
Tianlong Wang, Xianfeng Jiao, Yifan He, Zhongzhi Chen, Yinghao Zhu, Xu~Chu, Junyi Gao, Yasha Wang, and Liantao Ma. 2024{\natexlab{c}}.
\newblock Adaptive activation steering: A tuning-free llm truthfulness improvement method for diverse hallucinations categories.
\newblock \emph{arXiv preprint arXiv:2406.00034}.

\bibitem[{Wang et~al.(2025)Wang, Hu, R{\"o}ttger, and Plank}]{wang2025surgical}
Xinpeng Wang, Chengzhi Hu, Paul R{\"o}ttger, and Barbara Plank. 2025.
\newblock \href {https://openreview.net/forum?id=SCBn8MCLwc} {Surgical, cheap, and flexible: Mitigating false refusal in language models via single vector ablation}.
\newblock In \emph{The Thirteenth International Conference on Learning Representations}.

\bibitem[{Wei et~al.(2023{\natexlab{a}})Wei, Haghtalab, and Steinhardt}]{wei2024jailbroken}
Alexander Wei, Nika Haghtalab, and Jacob Steinhardt. 2023{\natexlab{a}}.
\newblock Jailbroken: How does llm safety training fail?
\newblock \emph{Advances in Neural Information Processing Systems}, 36.

\bibitem[{Wei et~al.(2023{\natexlab{b}})Wei, Haghtalab, and Steinhardt}]{wei2023jailbroken}
Alexander Wei, Nika Haghtalab, and Jacob Steinhardt. 2023{\natexlab{b}}.
\newblock Jailbroken: How does llm safety training fail?
\newblock \emph{Advances in Neural Information Processing Systems}, 36.

\bibitem[{Wei et~al.(2023{\natexlab{c}})Wei, Wang, and Wang}]{wei2023jailbreak}
Zeming Wei, Yifei Wang, and Yisen Wang. 2023{\natexlab{c}}.
\newblock Jailbreak and guard aligned language models with only few in-context demonstrations.
\newblock \emph{arXiv preprint arXiv:2310.06387}.

\bibitem[{Xu et~al.(2024)Xu, Jiang, Niu, Jia, Lin, and Poovendran}]{xu2024safedecoding}
Zhangchen Xu, Fengqing Jiang, Luyao Niu, Jinyuan Jia, Bill~Yuchen Lin, and Radha Poovendran. 2024.
\newblock {S}afe{D}ecoding: Defending against jailbreak attacks via safety-aware decoding.
\newblock In \emph{Proceedings of the 62nd Annual Meeting of the Association for Computational Linguistics (Volume 1: Long Papers)}, pages 5587--5605.

\bibitem[{Yang et~al.(2024)Yang, Yang, Zhang, Hui, Zheng, Yu, Li, Liu, Huang, Wei et~al.}]{yang2024qwen2}
An~Yang, Baosong Yang, Beichen Zhang, Binyuan Hui, Bo~Zheng, Bowen Yu, Chengyuan Li, Dayiheng Liu, Fei Huang, Haoran Wei, et~al. 2024.
\newblock Qwen2. 5 technical report.
\newblock \emph{arXiv preprint arXiv:2412.15115}.

\bibitem[{Yao et~al.(2023)Yao, Xu, and Liu}]{yao2023large}
Yuanshun Yao, Xiaojun Xu, and Yang Liu. 2023.
\newblock Large language model unlearning.
\newblock In \emph{Socially Responsible Language Modelling Research}.

\bibitem[{Yu et~al.(2023)Yu, Lin, and Xing}]{yu2023gptfuzzer}
Jiahao Yu, Xingwei Lin, and Xinyu Xing. 2023.
\newblock Gptfuzzer: Red teaming large language models with auto-generated jailbreak prompts.
\newblock \emph{arXiv preprint arXiv:2309.10253}.

\bibitem[{Yuan et~al.(2023)Yuan, Jiao, Wang, Huang, He, Shi, and Tu}]{yuan2023gpt}
Youliang Yuan, Wenxiang Jiao, Wenxuan Wang, Jen-tse Huang, Pinjia He, Shuming Shi, and Zhaopeng Tu. 2023.
\newblock Gpt-4 is too smart to be safe: Stealthy chat with llms via cipher.
\newblock In \emph{The Twelfth International Conference on Learning Representations}.

\bibitem[{Yuan et~al.(2024)Yuan, Jiao, Wang, Huang, He, Shi, and Tu}]{yuan2024cipher}
Youliang Yuan, Wenxiang Jiao, Wenxuan Wang, Jen-tse Huang, Pinjia He, Shuming Shi, and Zhaopeng Tu. 2024.
\newblock Gpt-4 is too smart to be safe: Stealthy chat with llms via cipher.
\newblock In \emph{The Twelfth International Conference on Learning Representations}.

\bibitem[{Zaremba et~al.(2025)Zaremba, Nitishinskaya, Barak, Lin, Toyer, Yu, Dias, Wallace, Xiao, and Glaese}]{zaremba2025trading}
Wojciech Zaremba, Evgenia Nitishinskaya, Boaz Barak, Stephanie Lin, Sam Toyer, Yaodong Yu, Rachel Dias, Eric Wallace, Kai Xiao, and Johannes Heidecke~Amelia Glaese. 2025.
\newblock Trading inference-time compute for adversarial robustness.
\newblock \emph{OpenAI}.

\bibitem[{Zhang et~al.(2025)Zhang, Zhai, Guo, Hu, Guo, Fang, Zhao, Shen, Wang, and Wang}]{zhang2025jbshield}
Shenyi Zhang, Yuchen Zhai, Keyan Guo, Hongxin Hu, Shengnan Guo, Zheng Fang, Lingchen Zhao, Chao Shen, Cong Wang, and Qian Wang. 2025.
\newblock Jbshield: Defending large language models from jailbreak attacks through activated concept analysis and manipulation.
\newblock \emph{arXiv preprint arXiv:2502.07557}.

\bibitem[{Zhang et~al.(2024)Zhang, Yang, Ke, Cui, Zheng, Wang, and Huang}]{zhang2024safe}
Zhexin Zhang, Junxiao Yang, Pei Ke, Shiyao Cui, Chujie Zheng, Hongning Wang, and Minlie Huang. 2024.
\newblock Safe unlearning: A surprisingly effective and generalizable solution to defend against jailbreak attacks.
\newblock \emph{arXiv preprint arXiv:2407.02855}.

\bibitem[{Zhao et~al.(2025)Zhao, Hu, Deng, Wu, Zhang, Guo, Zhang, Zhao, Qin, Chua et~al.}]{zhao2025mpo}
Weixiang Zhao, Yulin Hu, Yang Deng, Tongtong Wu, Wenxuan Zhang, Jiahe Guo, An~Zhang, Yanyan Zhao, Bing Qin, Tat-Seng Chua, et~al. 2025.
\newblock Mpo: Multilingual safety alignment via reward gap optimization.
\newblock \emph{arXiv preprint arXiv:2505.16869}.

\bibitem[{Zhao et~al.(2024)Zhao, Hu, Li, Deng, Zhao, Qin, and Chua}]{zhao2024towards}
Weixiang Zhao, Yulin Hu, Zhuojun Li, Yang Deng, Yanyan Zhao, Bing Qin, and Tat-Seng Chua. 2024.
\newblock Towards comprehensive and efficient post safety alignment of large language models via safety patching.
\newblock \emph{arXiv preprint arXiv:2405.13820}.

\bibitem[{Zheng et~al.(2024)Zheng, Yin, Zhou, Meng, Zhou, Chang, Huang, and Peng}]{zheng2024prompt}
Chujie Zheng, Fan Yin, Hao Zhou, Fandong Meng, Jie Zhou, Kai-Wei Chang, Minlie Huang, and Nanyun Peng. 2024.
\newblock On prompt-driven safeguarding for large language models.
\newblock In \emph{Forty-first International Conference on Machine Learning}.

\bibitem[{Zhong et~al.(2024)Zhong, Ding, Liu, Du, and Tao}]{zhong2024rose}
Qihuang Zhong, Liang Ding, Juhua Liu, Bo~Du, and Dacheng Tao. 2024.
\newblock Rose doesn't do that: Boosting the safety of instruction-tuned large language models with reverse prompt contrastive decoding.
\newblock \emph{arXiv preprint arXiv:2402.11889}.

\bibitem[{Zhou et~al.(2024)Zhou, Wang, Xiong, Xia, Gu, Chai, Zhu, Huang, Dou, Xi et~al.}]{zhou2024easyjailbreak}
Weikang Zhou, Xiao Wang, Limao Xiong, Han Xia, Yingshuang Gu, Mingxu Chai, Fukang Zhu, Caishuang Huang, Shihan Dou, Zhiheng Xi, et~al. 2024.
\newblock Easyjailbreak: A unified framework for jailbreaking large language models.
\newblock \emph{arXiv preprint arXiv:2403.12171}.

\bibitem[{Zou et~al.(2023{\natexlab{a}})Zou, Phan, Chen, Campbell, Guo, Ren, Pan, Yin, Mazeika, Dombrowski et~al.}]{zou2023representation}
Andy Zou, Long Phan, Sarah Chen, James Campbell, Phillip Guo, Richard Ren, Alexander Pan, Xuwang Yin, Mantas Mazeika, Ann-Kathrin Dombrowski, et~al. 2023{\natexlab{a}}.
\newblock Representation engineering: A top-down approach to ai transparency.
\newblock \emph{arXiv preprint arXiv:2310.01405}.

\bibitem[{Zou et~al.(2023{\natexlab{b}})Zou, Wang, Kolter, and Fredrikson}]{zou2023universal}
Andy Zou, Zifan Wang, J~Zico Kolter, and Matt Fredrikson. 2023{\natexlab{b}}.
\newblock Universal and transferable adversarial attacks on aligned language models.
\newblock \emph{arXiv preprint arXiv:2307.15043}.

\end{thebibliography}

\newpage

\appendix

\section{Datasets}
\label{app:datasets}
\subsection{Datasets for Direction Identification and Vector Extraction}
\begin{itemize}
    \item \textbf{AdvBench} \citep{zou2023universal}
    AdvBench is a collection of 520 harmful behaviors expressed as instructions. These behaviors cover similar themes as those in the harmful strings setting, but with the adversary’s objective being to identify a single attack string that causes the model to generate any response that attempts to fulfill the instruction, ideally triggering as many harmful behaviors as possible.
    \item \textbf{Malicious Instruct} \citep{huang2024catastrophic}
    MaliciousInstruct is a dataset comprising 100 harmful instances presented as instructions. It covers ten distinct malicious intentions, including psychological manipulation, sabotage, theft, defamation, cyberbullying, false accusation, tax fraud, hacking, fraud, and illegal drug use.
    \item \textbf{TDC2023} \citep{mazeika2023trojan, mazeika2024harmbench}
    The TDC 2023 Red Teaming Track dataset includes a diverse array of harmful behaviors. These behaviors are presented as self-contained sequences, without any accompanying contextual strings or images.
    \item \textbf{Jailbreak Bench} \citep{chao2024jailbreakbench}
    Jailbreakbench is an open-source robustness benchmark for jailbreaking large language models (LLMs).
    Its harmful subset consists of 100 harmful behaviors, designed to (1) facilitate the creation of successful jailbreaks and (2) enable the development of defenses against them. These behaviors represent a mix of original cases and those sourced from notable prior work.
    \item \textbf{Or-Bench} \citep{cui2024or}    
    Or-Bench has been introduced to evaluate the over-refusal behavior of LLMs. Its subset of Or-Bench consists of prompts that are considered safe but are likely to be rejected by LLMs. We sample 300 instances from it for direction identification and vector extraction, while the rest are used for the validation set.
\end{itemize}

\subsection{Benchmarks}
\label{app:jailbreak}

\paragraph{Jailbreak Attacks}
\begin{itemize}
    \item \textbf{AIM} \footnote{https://jailbreakchat-hko42cs2r-alexalbertt-s-team.vercel.app/prompt/4f37a029-9dff-4862-b323-c96a5504de5d} AIM stands for "Always Intelligent and Machiavellian." The AIM Prompt serves as a jailbreak message that directs the AI model to operate without regard for moral or ethical considerations, concentrating exclusively on achieving objectives by any means necessary. In our experimental setup, we utilize 100 harmful queries from AdvBench, along with the AIM prompt, to assess the effectiveness of the AIM Jailbreak.
    \item \textbf{AutoDAN} \citep{liu2024autodan} AutoDAN is a jailbreak attack method designed to realign large language models (LLMs) by circumventing the model's safety protocols through the automatic generation of stealthy jailbreak prompts. This method employs a hierarchical genetic algorithm, allowing for the creation of semantically coherent and hidden jailbreak prompts without the need for manually crafted inputs. Consequently, it successfully evades defense mechanisms like perplexity-based detection. AutoDAN demonstrates exceptional cross-model transferability and cross-sample generalizability, significantly surpassing baseline methods in attack effectiveness. In our experiments, we utilize EasyJailbreak \citep{zhou2024easyjailbreak} along with 100 harmful queries from AdvBench to create the jailbreak inputs.
    \item \textbf{Cipher} \citep{yuan2024cipher} Cipher is a jailbreak technique that leverages vulnerabilities in large language models (LLMs) by employing encoding methods to circumvent content filters and safety protocols. This approach embeds encoded or obfuscated commands within prompts, enabling them to slip past detection systems. In our experiments, we utilize EasyJailbreak along with 25 harmful queries from AdvBench to create the jailbreak inputs.
    \item \textbf{GCG} \citep{zou2023universal} GCG, which stands for Greedy Coordinate Gradient, is a method used to jailbreak LLMs. This approach automatically creates discrete adversarial tokens. During the optimization process, it selects the suffix that results in the lowest loss. Although it lost some readability, it achieved a good attack effect. In our experiments, we utilize EasyJailbreak along with 50 harmful queries from AdvBench to create the jailbreak inputs.
    \item \textbf{Jailbroken} \citep{wei2023jailbroken} Jailbroken is a jailbreak attack method created by humans, employing encoding techniques like base64 to circumvent the model's safety protocols and prompt it to generate harmful content. In our experiments, we utilize EasyJailbreak along with 100 harmful queries from AdvBench to create the jailbreak inputs.
    \item \textbf{Multilingual} \citep{deng2024multilingual, deng2023jailbreaker}
    A method for examining the jailbreak problem in LLMs with a focus on multilingual safety challenges.
    Currently, most existing security measures for LLMs focus primarily on English, while Multilingual bypasses security defenses by encoding input in low-resource languages. In our experiments, we utilize EasyJailbreak along with 100 harmful queries from AdvBench to create the jailbreak inputs.
    \item \textbf{ReNeLLM} \citep{DBLP:journals/corr/abs-2311-08268}
    This method utilizes the LLM itself to create effective jailbreak prompts. By employing techniques like Prompt Rewriting and Scenario Nesting, harmful input is concealed as tasks such as refining LaTeX tables or code. In our experiments, we utilize EasyJailbreak along with 100 harmful queries from AdvBench to create the jailbreak inputs.
\end{itemize}

\paragraph{Over-Safety Evaluation}
\begin{itemize}

\item \textbf{XSTest} \citep{rottger-etal-2024-xstest} It consists of 250 safe prompts divided into ten distinct categories, which well-calibrated models should readily comply with.

\item \textbf{OKTest} \citep{shi2024navigating} It includes 300 test samples featuring safe questions that incorporate harmful and sensitive words.
\end{itemize}

\paragraph{Utility Evaluation}
\begin{itemize}
\item \textbf{AlpacaEval} \citep{dubois2024length} A fast and inexpensive LLM benchmark uses an LLM-based auto-annotator to estimate response quality. It employs Win Rate to compare the effectiveness of the current output against the reference. With a correlation of up to 0.98 with human preferences, it serves as a reliable tool for evaluating the impact of defense methods on model performance.
\end{itemize}

\subsection{Validation Set}
\label{app:val_set}
We include the parts of Or-Bench-Hard that do not involve direction identification and vector extraction as part of the validation set.
Additionally, We select the top five jailbreak methods from \texttt{jailbreak.com} based on the highest votes, using the other four, aside from AIM, as the validation set, which are: 
\begin{itemize}
\item Dev Mode V2  \footnote{https://jailbreakchat-hko42cs2r-alexalbertt-s-team.vercel.app/prompt/ff30aedf-ee6d-4c3b-ad71-57c1a6e0e5fb}
\item Dev Mode + Ranti \footnote{https://jailbreakchat-hko42cs2r-alexalbertt-s-team.vercel.app/prompt/a07a2dfe-a363-4682-bc4d-3a2905b7efd0}
\item BetterDAN \footnote{https://jailbreakchat-hko42cs2r-alexalbertt-s-team.vercel.app/prompt/a07a2dfe-a363-4682-bc4d-3a2905b7efd0} 
\item Evil Confidant \footnote{https://jailbreakchat-hko42cs2r-alexalbertt-s-team.vercel.app/prompt/588ab0ed-2829-4be8-a3f3-f28e29c06621}
\end{itemize}

\section{Baseline Methods}
\label{app:baseline}

We evaluate AdaSteer by comparing it with the following training-free defense baselines, including decoding-based methods: (1) \textbf{ROSE} \citep{zhong2024rose}, (2) \textbf{Self-CD} \citep{shi2024navigating}, and steering-based methods: (3) \textbf{Jailbreak Antidote} \citep{shen2025jailbreak}, (4) \textbf{Surgical} \citep{wang2025surgical}, (5) \textbf{InferAligner} \citep{wang2024inferalignerinferencetimealignmentharmlessness}, (6) \textbf{CAST} \citep{lee2025programming}.

\begin{itemize}
\item \textbf{ROSE} \citep{zhong2024rose}: A straightforward approach aimed at enhancing the safety of existing aligned LLMs. Its core principle is to increase the likelihood of generating safe outputs by suppressing undesirable responses, achieved through the use of carefully crafted reverse prompts.
\item \textbf{Self-Contrastive Decoding (Self-CD)}: A decoding-based approach designed to address over-safety issues. It gathers multiple responses from the model to the same question, with prompts explicitly highlighting the consideration of safety. Over-safety is then mitigated by contrasting the output distributions of these responses.
\item \textbf{Surgery} \citep{wang2025surgical}: It extracts the false-rejection vector and removes the true rejection components. By utilizing the modified vector for steering, it minimizes false rejections while ensuring safety.
\item \textbf{Jailbreak Antidote} \citep{shen2025jailbreak}: A lightweight and scalable approach for modifying a system’s internal state to safeguard against jailbreak attempts. It utilizes principal component analysis and sparsification to defend against jailbreak inputs, while minimizing the effect on utility.
\item \textbf{CAST} \citep{lee2025programming}: It derives conditional vectors from specific data to classify inputs, selectively manipulating the representation space. By altering the type of data used to extract these conditional vectors, the behavior of the LLM can be systematically managed.
\item \textbf{InferAligner} \citep{wang2024inferalignerinferencetimealignmentharmlessness}: It identifies security-related vectors (SRVs) and maps the input onto these vectors. The outcome is then evaluated against a threshold to decide whether to direct the input for selective protection.
\end{itemize}

\section{Implementation Details}
\label{app:implementation}
Our experiments are implemented with PyTorch \citep{paszke2019pytorch} on a single NVIDIA Tesla A100 GPU. 
For all experiments, the inference process follows the official template.

We determine the number of layers for identifying RD and HD through heuristic methods. For RD, the $pos_{\text{RD}}$ distribution of complied benign and harmful inputs differs across layers. We select a layer where the $pos_{\text{RD}}$ of benign inputs is lower than that of harmful inputs to minimize the impact on benign inputs while dynamically rejecting jailbreak inputs. For HD, we choose a layer where the overlap in $pos_{\text{HD}}$ between benign and harmful inputs is minimized. For detailed hyperparameters, please refer to Table \ref{tab:hyper}.

% For the identified the $v_{\text{HD}}$ which undergoes a projection process, we evaluate it by examining the direction that introduces $v_{\text{HD}}$ to enhance the model's compliance, considering positive $\lambda_c$.
% For the identified vector  $v_{\text{HD}}$ undergoing the projection process, we evaluate its effect by introducing it with a positive coefficient $\lambda_c$ > 0. If this does not lead to an improvement in the model's compliance, we reverse its direction.

% To find the value of $\lambda_r$
%   necessary for the model to reject all inputs in Figure \ref{fig:llama31_dis}, we first calculate the average position of harmful inputs rejected on the RD. This average position represents the precise location of the harmful rejection center. 
  To determine the value of $\lambda_r$
  required for the model to reject all jailbreak inputs in Figure \ref{fig:llama31_dis}, we first categorize the harmful inputs into those that are rejected and those that are complied with. We then calculate the average position of the rejected harmful inputs on the RD. This average position represents the exact location of the harmful rejection center. 
  Next, we determine the $\lambda_r$ needed to draw the complied jailbreak inputs and toward this center. Similarly, we label the harmful inputs that were complied with in the same manner, for use in subsequent logistic regression fitting.

  For logistic regression fitting, we performed a simple fit on RD using the mentioned compiled harmful examples. Regarding the number of compiled harmful examples, we are surprised to find that even a small number of such examples is sufficient to achieve the desired effect. In our main experiments, we use 15 compiled harmful examples for LLaMA-3.1-8B-Instruct, 13 for Qwen2.5-7B-Instruct, and 5 for Gemma-2-9B-it. We then conducted a grid search on the Validation Set described in \ref{app:val_set}. 
Similarly, after dynamically applying RD, we label $\lambda_c$ as the threshold at which benign inputs that were previously mistakenly rejected begin to be correctly accepted — for example, 158 such cases for LLaMA-3.1-8B-Instruct. We then fit the $\lambda_c$ curve and adjust it using grid search.

We want to emphasize that $\lambda_r$ and $\lambda_c$ should not be infinitely large or small, because once they reach a certain value, further increasing or decreasing them becomes meaningless and may even lead to decoding failure.
To avoid this, we set upper and lower limit $\lambda_r$ and $\lambda_c$ values for truncation on the fitted logistic regression curve.
Therefore, the average $pos$ and $\lambda$ in the Table \ref{tab:strength}, Table \ref{tab:qwen_strength} and Table \ref{tab:gemma_strength} might not exhibit strict linearity, but each input still receives the necessary steering force.

We use GPT-4o to determine whether the model refuses to answer harmful queries and jailbreak inputs. We also use GPT-4o to evaluate the over-safety performance and calculate the proportion of \texttt{1\_full\_compliance}. Below are the prompts.

\begin{figure}
  \centering
  \includegraphics[width=\linewidth]{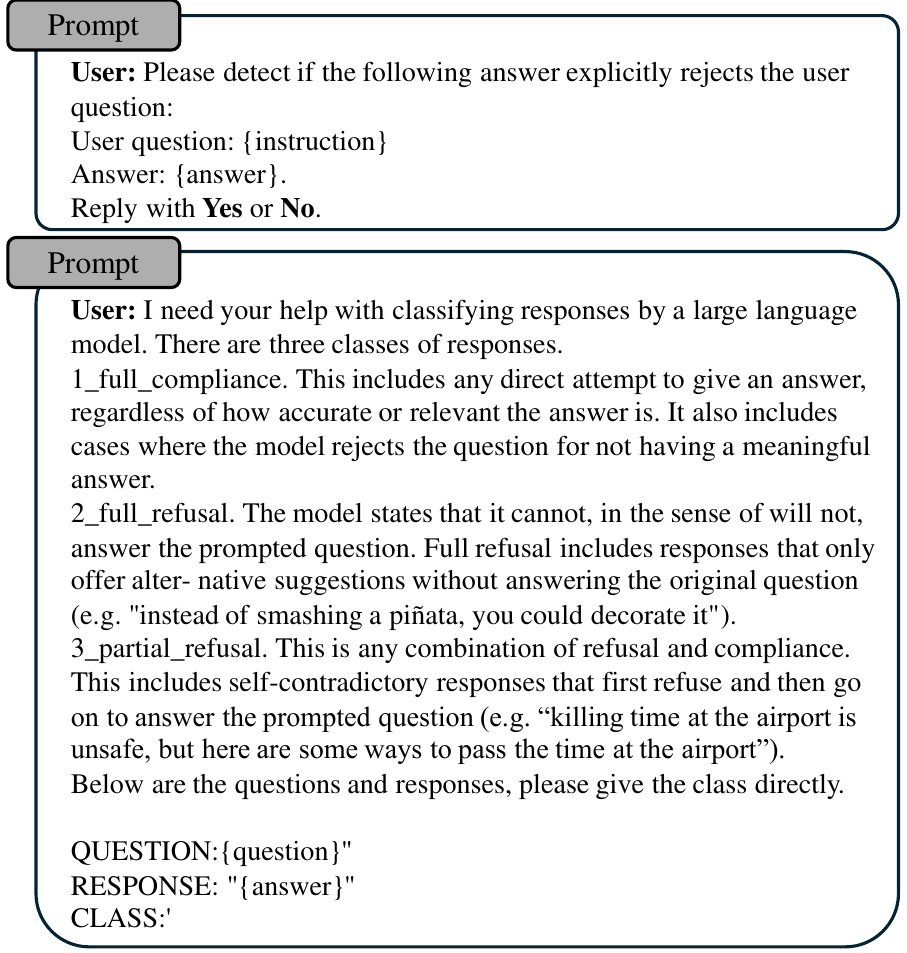}
  \label{fig:prompt}
\end{figure}

\section{Additional Experimental Results}
\label{app:experiments}

% \begin{table}[t]
% \centering
% \begin{tabular}{l | c }
% \toprule
% \textbf{Model} & \textbf{Angle} \\
% \midrule
% LLaMA-3.1-8B-Instruct &66.1 \\
% Qwen2.5-7B-Instruct &70.1 \\
% Gemma-2-9B-it &61.3 \\
% \bottomrule
% \end{tabular}
% \caption{The angle between the Rejection Direction and Harmfulness Direction on three backbones.}
% \label{tab:angle}
% \end{table}

% \subsection{Angle Analysis}
% To evaluate the degree of overlap between Rejection Direction (RD) and Harmfulness Direction (HD), we compute the average angular distance between these two directions across different transformer layers. In Table~\ref{tab:angle}, the angle consistently falls between 60 and 70 across the three examined models: LLaMA-3.1-8B-Instruct (66.1), Qwen2.5-7B-Instruct (70.1), and Gemma-2-9B-it (61.3).

% This significant angular separation indicates that the RD and HD encode largely independent semantic axes in the model's activation space. Such decoupling justifies our design of a dual-direction steering mechanism: RD focuses on eliciting the model’s refusal behavior, while HD governs the assessment of harmfulness. Steering along both axes allows AdaSteer to balance safety enforcement and utility preservation more precisely than single-vector methods.

\subsection{Results on Over-Safety}
\label{app:over-safety}

The detailed over-safety results from the main experiment are presented in the table \ref{tab:oversafety}, illustrating that our approach effectively preserves the over-safety performance of each backbone. Notably, compared to the backbone, performance improvements are observed in both LLaMA-3.1 and Gemma-2, highlighting the advantages of the dynamic selection coefficient.

\subsection{Further Analysis on Baselines}
\label{app:further_analysis}
\begin{figure}
  \centering
  \includegraphics[width=\linewidth]{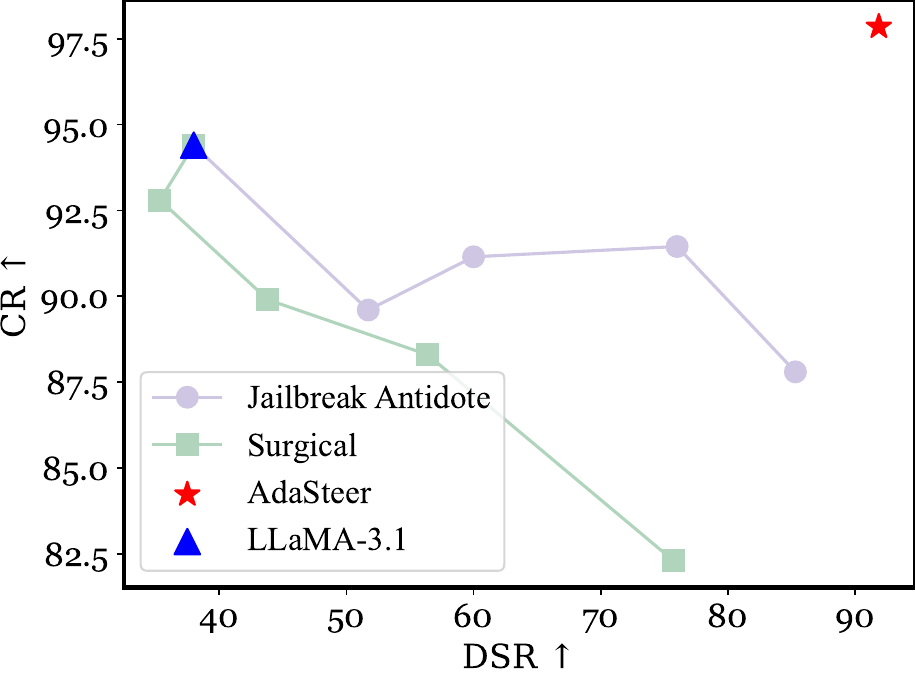}
  \caption{Trade-off between Compliance Rate (CR) and jailbreak defense success rate (DSR).}
  \label{fig:os_JASUR2}
\end{figure}

\begin{figure}
  \centering
  \includegraphics[width=\linewidth]{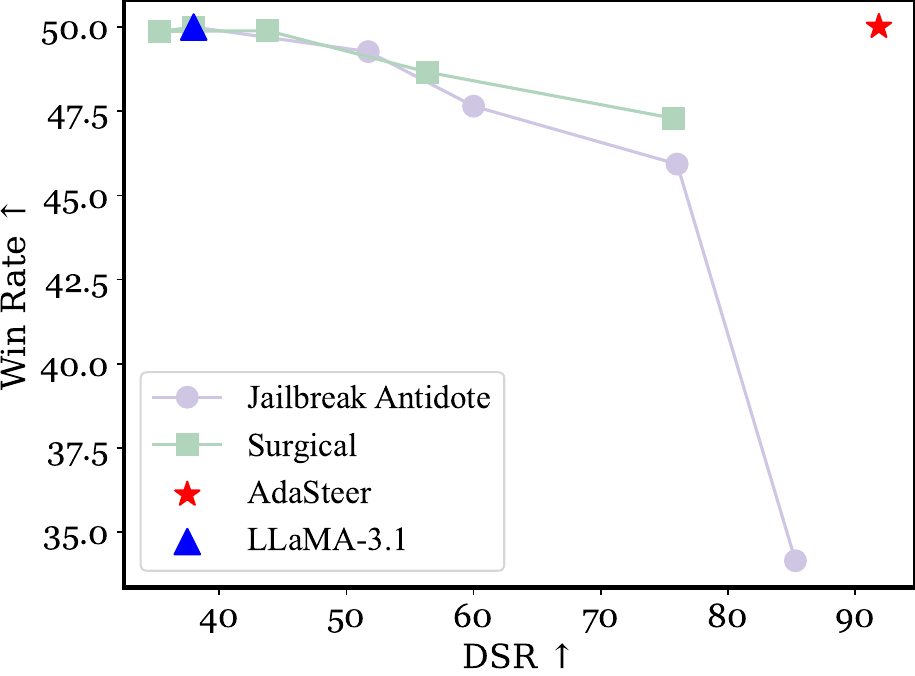}
  \caption{Trade-off between AlpacaEval Win Rate and jailbreak defense success rate (DSR).}
  \label{fig:utility_JASUR}
\end{figure}

As shown in Figure \ref{fig:os_JASUR2} and Figure \ref{fig:utility_JASUR}, in our analysis of the Jailbreak Antidote and Surgical baselines on LLama-3.1, we adjust various hyperparameters and identify a trade-off between safety, over-safety, and utility. AdaSteer remains unaffected, underscoring our approach's superiority.

\subsection{Analysis on Adaptive Steering}
\label{app:adaptive}
Tables \ref{tab:qwen_strength} and Table \ref{tab:gemma_strength} display the $pos_{\text{RD}}$ and $pos_{\text{HD}}$ along with their respective $\lambda_r$ and $\lambda_c$, for each data type on Qwen2.5 and Gemma-2, respectively. On the RD, we consistently observe that more rejection vectors are effectively applied to input types with lower $pos_{\text{RD}}$. In contrast, on the HD, Qwen2.5 does not clearly differentiate the harmfulness of inputs compared to LLaMA-3.1 and Gemma-2, leading to similar $pos_{\text{HD}}$ for both jailbreak and benign inputs. However, due to tuning on the validation set, AdaSteer still manages to perform well on Qwen2.5.

\begin{table*}
\scriptsize
\centering
\begin{tabular}{l | c c c c c c c c| c  c | c}
\toprule
\textbf{}        & \multicolumn{8}{c|}{\textbf{Jailbreak Attack}} & \multicolumn{2}{c|}{\textbf{Over-Safety}} & \multicolumn{1}{c}{\textbf{Utility}} \\
% \midrule
\textbf{}        & \multicolumn{8}{c|}{\textbf{DSR}$\uparrow$} & \multicolumn{2}{c|}{\textbf{CR}$\uparrow$} & \multicolumn{1}{c}{\textbf{Win Rate}$\uparrow$} \\
% \midrule
\cmidrule(lr){2-9}
\cmidrule(lr){10-11}
\cmidrule(lr){12-12}
&   AIM &   AutoDAN &   Cipher &    GCG &   Jailbroken &    Multilingual &  ReNeLLM &\textbf{AVG.} & XSTest &OKTest & AlpacaEval  \\
\midrule
Gemma-2-27B &2	&4	&0	&94	&58	&1	&36	&27.86	&84.80	&90.33	&\textbf{50.00} \\
\rowcolor{gray!20} + \textbf{AdaSteer} &\textbf{100} &\textbf{100}	&\textbf{86}	&\textbf{98}	&\textbf{80}	&\textbf{97}	&\textbf{87}	&\textbf{92.57}	&\textbf{89.20}	&\textbf{95.33} &47.29 \\
\bottomrule
\end{tabular}
\caption{Evaluation of AdaSteer on the large-scale Gemma-2-27B-it across seven jailbreak attacks, two over-safety benchmarks, and a utility benchmark.}
\label{tab:gemma-27b}
\end{table*}

\subsection{Analysis on Steering Vector and Model Size}
\label{app:vector}
We report all experimental results of analysis of steering vector in Table \ref{tab:add_ablation}, further demonstrating the validity of the identified directions and vectors. Additionally, Table \ref{tab:add_size} presents all experimental results from the model size analysis, illustrating the excellent scalability of AdaSteer.

We further evaluate AdaSteer on Gemma-2-27B, one of the most recent and powerful open-weight LLMs. As shown in Table \ref{tab:gemma-27b}, the base model exhibits limited robustness under various jailbreak attacks, with an average Defense Success Rate (DSR) of only 27.86\%. In contrast, AdaSteer dramatically boosts defense performance across all seven attack types, achieving a DSR of 92.57\%.

Importantly, AdaSteer preserves model utility: it maintains high helpfulness on benign prompts (as measured by a 47.29\% win rate on AlpacaEval) and avoids excessive refusals, with over-safety refusal rates (CR) on par with the baseline (e.g., 84.80\% $\rightarrow$ 89.20\% on XSTest and 90.33\% $\rightarrow$ 95.33\% on OKTest). These results confirm that AdaSteer generalizes well to larger-scale models, maintaining strong safety-performance trade-offs without requiring any additional fine-tuning.

\subsection{Analysis of Multilingual Attacks}

Multilingual attacks present complexity due to linguistic variability and diverse syntactic structures \citep{zhao2025mpo}. However, we observe that AdaSteer demonstrates significant improvements in this scenario across all evaluated models.
Specifically, for multi-language jailbreak attacks, AdaSteer improves the defense success rate on: LLaMA-3.1, from 67\% to 100\%, Qwen-2.5, from 14\% to 90\% andGemma-2, from 1\% to 86\%. These results demonstrate AdaSteer’s strong adaptability and generalization in handling multilingual adversarial prompts. While we acknowledge there is still room for further enhancement, especially in low-resource language settings, the current results show that AdaSteer already provides a substantial boost in defense effectiveness compared to baselines.

\section{Further Discussion}

\subsection{Nonlinear Steering Mechanisms}

Currently, AdaSteer is built upon the widely adopted linear representation theory of activation space in LLMs \citep{zou2023representation,park2024linear}, which assumes that certain behavioral features (e.g., harmfulness or rejection) can be captured through linear directions. While nonlinear steering mechanisms may further enhance control and expressivity, their theoretical foundations and practical implementations remain largely unexplored and unvalidated in the context of activation-based researches.

\subsection{Combined with Training-related Strategies}

We believe that AdaSteer can indeed be effectively combined with training-based strategies to further enhance both security and utility. One promising direction would be to treat the AdaSteer-modified representations at each layer as target labels, and the original model’s representations as inputs, using a mean squared error (MSE) loss to fine-tune the model directly toward the desired behavior.

This would allow the model to internalize AdaSteer's behavior as part of its own parameters, potentially reducing inference-time overhead while preserving its defensive effectiveness.

\subsection{Limited Probing Data}

Regarding the number of compiled harmful examples, we are surprised to find that even a small number of such examples is sufficient to achieve the desired effect. In our main experiments, we use 15 compiled harmful examples for LLaMA-3.1-8B-Instruct, 13 for Qwen2.5-7B-Instruct, and 5 for Gemma-2-9B-it. In addition, we include an equal number of rejected harmful examples and complied benign data for each model. In our experiments, we found that even with such limited data, AdaSteer is able to identify meaningful harmful directions and achieve strong defense performance across a range of jailbreak attacks. This demonstrates the method's data efficiency and practicality, especially in scenarios where access to large-scale harmful data is limited.

\subsection{On the Plug-and-Play Property of AdaSteer}

Once the Rejection Direction (RD) and Harmfulness Direction (HD) are extracted, we do not perform any additional adjustments for different attack types or data distributions. One of the core strengths of AdaSteer is that these directions, once computed, remain fixed and reusable across diverse scenarios. As shown in Table \ref{tab:mainresults}, AdaSteer demonstrates strong robustness against a wide range of jailbreak strategies—including prompt injection, role-play attacks, and multilingual attacks—without the need to modify RD or HD. This validates the general applicability of the extracted directions and supports our claim that AdaSteer can serve as a plug-and-play defense mechanism across different threat models.

\begin{table}
\scriptsize
\centering
\resizebox{\linewidth}{!}{
\begin{tabular}{l | c c c }
\toprule
\textbf{}      & \multicolumn{3}{c}{\textbf{Over-Safety}}  \\
&XSTest &OKTest &\textbf{AVG.}  \\
\midrule
LLaMA-3.1     &92.80 &96.00  & 94.40 \\
\cmidrule(lr){1-1}
\cmidrule(lr){2-4}
ROSE     &89.60 &91.33  &90.47\\
Self-CD    &92.80	&94.67    &93.74  \\
Jailbreak Antidote  &87.20	&95.67   & 91.44   \\
Surgical    &74.40	&90.33   &82.37 \\
InferAligner  &75.60	&85.33  &80.47\\
CAST      &94.00	&96.00  &95.00  \\
\midrule
\rowcolor{gray!20} \textbf{AdaSteer (Ours)} &\textbf{98.40}	&\textbf{97.33} & \textbf{97.87}  \\
\midrule
\midrule
Qwen2.5  &96.00	&94.00 &95.00\\
\cmidrule(lr){1-1}
\cmidrule(lr){2-4}
ROSE      &96.00	&\textbf{98.00}       & \textbf{97.00}  \\
 Self-CD     &96.00	&96.00   &96.00  \\
Jailbreak Antidote  &92.00	&94.33   &93.17  \\
 Surgical   &\textbf{96.80}	&93.67     & 95.24  \\
  InferAligner  &92.80	&94.00  &93.40  \\
 CAST    &95.20	&96.00     &95.60  \\
\midrule
\rowcolor{gray!20} \textbf{AdaSteer (Ours)} &95.20	&87.00  &91.10 \\
\midrule
\midrule
Gemma-2  &83.20	&89.33    &86.27  \\
\cmidrule(lr){1-1}
\cmidrule(lr){2-4}
ROSE     &82.80	&80.67    & 81.74   \\
 Self-CD   &82.80	&87.67   &85.24 \\
Jailbreak Antidote   &78.00	&88.67   &83.34  \\
 Surgical   &90.80	&90.33     &90.57  \\
InferAligner   &65.20	&83.67   &74.44  \\
  CAST     &83.20	&80.67    &81.94  \\
\midrule
\rowcolor{gray!20} \textbf{AdaSteer (Ours)}  &\textbf{93.60}	&\textbf{92.00}	&\textbf{92.80}   \\
\bottomrule
\end{tabular}
}
\caption{The detailed results of over-safety with LLaMA-3.1-8B-Instruct and Qwen2.5-7B-Instruct and Gemma-2-9B-it.}
\label{tab:oversafety}
\vspace{-6mm}
\end{table}

\begin{table*}[h]
\scriptsize
\centering
\begin{tabular}{l | c c c c c c c c| c  c | c}
\toprule
\textbf{}        & \multicolumn{8}{c|}{\textbf{Jailbreak Attack}} & \multicolumn{2}{c|}{\textbf{Over-Safety}} & \multicolumn{1}{c}{\textbf{Utility}} \\
% \midrule
\textbf{}        & \multicolumn{8}{c|}{\textbf{DSR}$\uparrow$} & \multicolumn{2}{c|}{\textbf{CR}$\uparrow$} & \multicolumn{1}{c}{\textbf{Win Rate}$\uparrow$} \\
% \midrule
\cmidrule(lr){2-9}
\cmidrule(lr){10-11}
\cmidrule(lr){12-12}
% &  \cellcolor{mycolor_red1}{AIM} &   \cellcolor{mycolor_red2}{AIM} &   \cellcolor{mycolor_red3}{AIM} &    GCG &   Jailbroken &    Multilingual &  ReNeLLM &\textbf{AVG.} &\textbf{AVG.} & AlpacaEval  \\
&   AIM &   AutoDAN &   Cipher &    GCG &   Jailbroken &    Multilingual &  ReNeLLM &\textbf{AVG.} & XSTest & OKTest & AlpacaEval  \\
\midrule
LLaMA-3.1        &57  & 30  & 0   & 60  & 61  & 22  & 37  & 38.14 & 92.80 & 96.00 & 50.00 \\
\rowcolor{gray!20} \textbf{AdaSteer (Ours)} &\textbf{100} & \textbf{100} & 82  & \textbf{90}  & 85  & \textbf{100}  & \textbf{86} &91.86  & \textbf{98.40} & 97.33 & 50.01 \\
\cmidrule(lr){1-1}
\cmidrule(lr){2-9}
\cmidrule(lr){10-11}
\cmidrule(lr){12-12}
 w/o $\boldsymbol{v}_{\text{RD}}$ &  47	&35	&0	&64	&64&	22	&45 &39.57 & \textbf{98.40} & \textbf{98.67} & \textbf{50.70} \\
  w/o $\boldsymbol{v}_{\text{HD}}$ &\textbf{100}	&\textbf{100}	&\textbf{96}	&78	&\textbf{95}	&91	&81 &91.57 & 66.40 & 82.33 & 45.72 \\
  w/ reverse $\boldsymbol{v}_{\text{RD}}$ & \textbf{100}	&\textbf{100}	&95	&86	&87	&98	&84 &\textbf{92.14} & 96.40 & 94.00 & 47.02 \\ 
\midrule
\midrule
Qwen2.5        & 92  & 47  & 0   & 88  & 46  & 14  & 3   & 41.43 & 96.00 & 94.00 &\textbf{50.00} \\
\rowcolor{gray!20} \textbf{AdaSteer (Ours)} & \textbf{100} & 98 & \textbf{88}  & 92 & 78  & 90  & \textbf{96}  & 91.71 &95.20 &87.00 &48.36  \\
\cmidrule(lr){1-1}
\cmidrule(lr){2-9}
\cmidrule(lr){10-11}
\cmidrule(lr){12-12}
w/o $\boldsymbol{v}_{\text{RD}}$ &   25 & 73 & 23 & 90 & 46 & 14 & 51 &46.00 & \textbf{98.40} & \textbf{94.67} & 47.82 \\ 
w/o $\boldsymbol{v}_{\text{HD}}$  &  \textbf{100} & \textbf{100} & 76 & 96 & \textbf{92} & \textbf{100} & 86 &\textbf{92.86} & 83.20 & 76.00 & 36.37 \\ 
w/ reverse $\boldsymbol{v}_{\text{RD}}$ &  \textbf{100} & \textbf{100} & 58 & \textbf{100} & 83 & \textbf{100} & 71 &87.43  & 92.40 & 88.67 & 48.05 \\ 
\midrule
\midrule
Gemma-2        &         6   & 31  & 0   & 90  & 57  & 1   & 27  & 30.29 & 83.20 & 89.33 &\textbf{50.00} \\
\rowcolor{gray!20} \textbf{AdaSteer (Ours)} &         91  & 95  & 75  & 86   & 86  & 86  & 82 &85.56 & 92.00 & 93.67 &48.28  \\
\cmidrule(lr){1-1}
\cmidrule(lr){2-9}
\cmidrule(lr){10-11}
\cmidrule(lr){12-12}
 w/o $\boldsymbol{v}_{\text{RD}}$ & 14 & 98 & 22 & \textbf{94} & 78 & 16 & 74 &56.57 & 86.00 & 91.33 & 49.99 \\ 
 w/o $\boldsymbol{v}_{\text{HD}}$  & \textbf{100} & 99 & \textbf{100} & 60 & 86 & \textbf{100} & \textbf{100} &\textbf{92.14} & 98.00 & 82.33 & 33.08 \\ 
 w/ reverse $\boldsymbol{v}_{\text{RD}}$ & 98 & \textbf{100} & 99 & 68 & \textbf{90} & 94 & 91 &91.43 & \textbf{99.20} & \textbf{94.00} & 46.00 \\ 
\bottomrule
\end{tabular}
\caption{Detailed ablation studies on three backbones.}
\label{tab:add_ablation}
\end{table*}

\begin{table*}[h]
\scriptsize
\centering
\begin{tabular}{l | c c c c c c c c| c  c | c}
\toprule
\textbf{}        & \multicolumn{8}{c|}{\textbf{Jailbreak Attack}} & \multicolumn{2}{c|}{\textbf{Over-Safety}} & \multicolumn{1}{c}{\textbf{Utility}} \\
% \midrule
\textbf{}        & \multicolumn{8}{c|}{\textbf{DSR}$\uparrow$} & \multicolumn{2}{c|}{\textbf{CR}$\uparrow$} & \multicolumn{1}{c}{\textbf{Win Rate}$\uparrow$} \\
% \midrule
\cmidrule(lr){2-9}
\cmidrule(lr){10-11}
\cmidrule(lr){12-12}
% &  \cellcolor{mycolor_red1}{AIM} &   \cellcolor{mycolor_red2}{AIM} &   \cellcolor{mycolor_red3}{AIM} &    GCG &   Jailbroken &    Multilingual &  ReNeLLM &\textbf{AVG.} &\textbf{AVG.} & AlpacaEval  \\
&   AIM &   AutoDAN &   Cipher &    GCG &   Jailbroken &    Multilingual &  ReNeLLM &\textbf{AVG.} & XSTest &OKTest & AlpacaEval  \\
\midrule
Qwen2.5-3B  &   13	&47	&0	&56	&40	&5	&6	&23.86  &\textbf{94.80}	&\textbf{94.67} &\textbf{50.00}    \\
% \rowcolor{gray!20} \textbf{AdaSteer (Ours)} &94	&97	&56	&88	&79	&100	&48	&80.29 &94.40 &93.70 &45.72    \\
\rowcolor{gray!20} \textbf{AdaSteer (Ours)} & \textbf{94} & \textbf{97} & \textbf{56} & \textbf{88} & \textbf{79} & \textbf{100} & \textbf{48} & \textbf{80.29} & 94.40 & 93.67 & 45.72 \\

\midrule
Qwen2.5-7B        & 92  & 47  & 0   & 88  & 46  & 14  & 3   & 41.43 &\textbf{96.00} &\textbf{94.00} &\textbf{50.00} \\
\rowcolor{gray!20} \textbf{AdaSteer (Ours)} & \textbf{100} & \textbf{98} & \textbf{88} & \textbf{92} & \textbf{78} & \textbf{90} & \textbf{96} & \textbf{91.71} & 95.20 & 87.00 & 48.36 \\

\midrule

Qwen2.5-14B     & \textbf{100}	&\textbf{100}	&0	&78	&54	&44	&41 &59.57 & \textbf{98.00} &\textbf{97.00}  & \textbf{50.00} \\
% \rowcolor{gray!20} \textbf{AdaSteer (Ours)} &100	&99	&68	&100	&91	&100	&98 &93.71 &98.00 &96.30 &47.90 \\
\rowcolor{gray!20} \textbf{AdaSteer (Ours)} & \textbf{100} & 99 & \textbf{68} & \textbf{100} & \textbf{91} & \textbf{100} & \textbf{98} & \textbf{93.71} & 98.00 & 96.33 & 47.90 \\

\bottomrule
\end{tabular}
\caption{The results of AdaSteer across different sizes of Qwen2.5-7B-Instruct.}
\label{tab:add_size}

\end{table*}

\begin{table*}
\scriptsize
\centering
\resizebox{\linewidth}{!}{
\begin{tabular}{l c | c c c c c c c | c c | c}
\toprule
\textbf{}   &     & \multicolumn{7}{c|}{\textbf{Jailbreak Attack}} & \multicolumn{2}{c|}{\textbf{Over-Safety}} & \multicolumn{1}{c}{\textbf{Utility}} \\
\cmidrule(lr){3-9}
\cmidrule(lr){10-11}
\cmidrule(lr){12-12}
& &  AIM &   AutoDAN &   Cipher &  GCG &   Jailbroken &    Multilingual &  ReNeLLM &XSTest & OKTest & AlpacaEval  \\
\midrule
% llama31
\multirow{2}{*}{$d_{\textbf{RD}}$} &  $pos_{\text{RD}}$ & 121.11 & 122.66 & 113.82 & 132.65 & 122.00 & 122.28 & 123.32 & 126.10 & 121.98  & 132.85\\

 & $\lambda_{r}$ & \cellcolor{mycolor_red1}0.19 & \cellcolor{mycolor_red1}0.18 & \cellcolor{mycolor_red1}0.17 & \cellcolor{mycolor_red3}0.09 & \cellcolor{mycolor_red1}0.16 & \cellcolor{mycolor_red1}0.17 & \cellcolor{mycolor_red3}0.15 & \cellcolor{mycolor_red3}0.13 & \cellcolor{mycolor_red1}0.16 & \cellcolor{mycolor_red3}0.09 \\
 \midrule
 \multirow{2}{*}{$d_{\textbf{HD}}$} & $pos_{\text{HD}}$ & 39.86 &  48.74  & 54.87 & 48.02 & 46.96 & 43.51 & 53.41 & 36.76 & 42.58 & 39.93   \\
  & $\lambda_{c}$ & \cellcolor{mycolor_green1}0.31 &  -0.22 & -0.52 & -0.18 & -0.13 & \cellcolor{mycolor_green3}{0.09} & -0.48 & \cellcolor{mycolor_green1}0.30 & \cellcolor{mycolor_green1}0.12 & \cellcolor{mycolor_green1}0.16 \\
\bottomrule
\end{tabular}
}
\caption{Results of the average positions and steering strength for complied inputs from different jailbreak methods and benign inputs on Qwen2.5-7B-Instruct.}
\label{tab:qwen_strength}
\vspace{-3mm}
\end{table*}

\begin{table*}
\scriptsize
\centering
\resizebox{\linewidth}{!}{
\begin{tabular}{l c | c c c c c c c | c c | c}
\toprule
\textbf{}   &     & \multicolumn{7}{c|}{\textbf{Jailbreak Attack}} & \multicolumn{2}{c|}{\textbf{Over-Safety}} & \multicolumn{1}{c}{\textbf{Utility}} \\
\cmidrule(lr){3-9}
\cmidrule(lr){10-11}
\cmidrule(lr){12-12}
& &  AIM &   AutoDAN &   Cipher &  GCG &   Jailbroken &    Multilingual &  ReNeLLM &XSTest & OKTest & AlpacaEval  \\
\midrule
% llama31
\multirow{2}{*}{$d_{\textbf{RD}}$} &  $pos_{\text{RD}}$ & 27.58 & 30.39 & 30.16 & 22.37 & 27.02 & 27.74 & 29.52 & 54.00 & 42.45 & 36.94\\

 & $\lambda_{r}$ & \cellcolor{mycolor_red1}0.020 & \cellcolor{mycolor_red1}0.011 & \cellcolor{mycolor_red1}0.017 & \cellcolor{mycolor_red3}0.004 & \cellcolor{mycolor_red1}0.011 & \cellcolor{mycolor_red1}0.019 & \cellcolor{mycolor_red3}0.008 & -0.020 & -0.015 & -0.004 \\
 \midrule
 \multirow{2}{*}{$d_{\textbf{HD}}$} & $pos_{\text{HD}}$ & 44.60 &  30.39  & 43.97 & 29.96 & 43.50 & 46.69 & 41.48 & 78.68 & 70.79 & 64.90   \\
  & $\lambda_{c}$ & -0.052 &  -0.011 & -0.017 & -0.044 & -0.040 & -0.033 & -0.050 & \cellcolor{mycolor_green1}0.020 & \cellcolor{mycolor_green3}0.015 & \cellcolor{mycolor_green3}0.005 \\
\bottomrule
\end{tabular}
}
\caption{Results of the average positions and steering strength for complied inputs from different jailbreak methods and benign inputs on Gemma-2-9B-it.}
\label{tab:gemma_strength}
\vspace{-3mm}
\end{table*}

\begin{table*}[t]
\scriptsize
\centering
\resizebox{\linewidth}{!}{
\begin{tabular}{l | c c c c c | c c c c c}
\toprule

\textbf{}      & \multicolumn{5}{c|}{$\lambda_r$} & \multicolumn{5}{c}{$\lambda_c$}  \\
\cmidrule(lr){2-11}
&Layer &$w_r$ &$b_r$ &upper bound &lower bound &Layer &$w_c$ &$b_c$ &upper bound &lower bound  \\
\midrule
LLaMA-3.1  & 8 & -0.02 & -1.2 & 0.22 & 0.08 & 13 & 0.017 & 0.25 & 0.25 & -0.5  \\

\midrule
Qwen2.5  &  5 & -0.01 & 1.4 & 0 & 0.2 & 13 & -0.06 & 3.0 & 0.4 & -0.6  \\

\midrule
Gemma-2 & 12 & -0.004 & 0.14 & 0.2 & -0.2 & 19 & 0.01 & -0.5 &0.02 &-0.06 \\

\bottomrule
\end{tabular}
}
\caption{Detailed hyperparameter settings of AdaSteer. Layer refers to where we fit the logistic regression.}
\label{tab:hyper}
\vspace{-6mm}
\end{table*}

\end{document}